\documentclass[11pt]{article}
\usepackage{mystyle}

\begin{document}

\title{\fontsize{20pt}{24pt}\selectfont Pseudodeterminism and $\MA\ne\NP^\BPP$ in Communication Complexity}
\author{\Large Thomas Watson\footnote{Department of Computer Science, University of Memphis. Supported by NSF grant CCF-1942742.}}
\maketitle

\begin{abstract}
We prove an exponential separation between zero-sided-error randomized and two-sided-error pseudodeterministic communication complexities of partial boolean functions. This qualitatively improves and simplifies the proof of the separation by G{\"{o}}{\"{o}}s, Harms, Riazanov, Sofronova, Sokolov, and Yuan (STOC 2026), which had a two-sided-error randomized upper bound. Then, we generalize our technique to separate the communication complexity analogues of $\MA$ and $\NP^\BPP$.
\end{abstract}


\section{Introduction} \label{sec:intro}

A randomized algorithm for a search problem (relation) is such that for every input, with high probability, the algorithm outputs a solution. A \emph{pseudodeterministic} algorithm for a search problem is a randomized algorithm such that for every input, there exists a solution $s$ such that with high probability, the algorithm outputs $s$. Thus ``pseudodeterministic'' is sandwiched between ``deterministic'' and ``randomized.'' One benefit of pseudodeterministic algorithms is that the success probability can be amplified, whereas for randomized algorithms it generally can't be (unless there's an efficient algorithm to check whether a given $s$ is indeed a solution). Pseudodeterminism has applications in cryptography, machine learning, and distributed computing, and has been studied in the settings of time complexity \cite{gat11probabilistic,goldreich13possibilities,oliveira17pseudodeterministic,dixon18pseudodeterministic,oliveira18pseudo,goldreich25multi,dixon21complete,lu21pseudodeterministic,dixon22pseudodeterminism,chen26polynomial}, space complexity \cite{grossman19reproducibility,goldwasser20pseudo,braverman23lower,grossman23tight}, query complexity \cite{goldreich13possibilities,goldwasser21pseudo,chattopadhyay25pseudo,gavinsky25unambiguous}, communication complexity \cite{huynh12virtue,blondal26borsuk,goos26pseudodeterministic}, parallel time complexity \cite{goldwasser17bipartite,ghosh21matroid}, interactive proofs \cite{goldwasser18pseudo,goemans19doubly}, and quantum computing \cite{aaronson26pseudo}.

Suppose $F\colon\{0,1\}^N\to\{0,1,\bot\}$ is a partial boolean function (promise decision problem), where $\bot$ means ``input is invalid; don't care about the output.'' We view $F$ as a search problem where $\bot$ means ``both $0$ and $1$ are solutions.'' On an invalid input, a pseudodeterministic algorithm must accept with either high or low probability (it doesn't matter which), whereas a randomized algorithm may accept with any probability. Thus, in any model of computation, the pseudodeterministic complexity of $F$ is the minimum randomized complexity of any total function $F'\colon\{0,1\}^N\to\{0,1\}$ that agrees with $F$.

We consider the model of two-party communication protocols. $\P\subseteq\ZPP\subseteq\RP\subseteq\BPP$ are the complexity classes of all (infinite families of) functions $F\colon\{0,1\}^N\times\{0,1\}^N\to\{0,1,\bot\}$ with $\polylog N$ deterministic, zero-sided-error, one-sided-error, and two-sided-error randomized communication complexity, respectively. (In the literature, sometimes a $^\cc$ superscript differentiates such classes from their time complexity counterparts.) Recall $\ZPP=\RP\cap\coRP$. Let $\psP$ be the analogous class for two-sided-error pseudodeterministic communication complexity. \cite{goos26pseudodeterministic} proved $\BPP\not\subseteq\psP$ by showing that some two-party partial boolean function has two-sided-error randomized communication complexity $O(\log N)$ and two-sided-error pseudodeterministic communication complexity $\widetilde{\Omega}(\sqrt{N})$. Their argument does not yield an upper bound with one-sided error. We improve the upper bound all the way to zero-sided error. Our result qualitatively improves, quantitatively matches, and somewhat simplifies the proof of the main result in \cite{goos26pseudodeterministic}.

\begin{theorem} \label{thm:zpp-psp}
$\ZPP\not\subseteq\psP$ in communication complexity.
\end{theorem}

It may appear unrelated, but this theorem is a stepping stone and warm-up for our next theorem.

Communication complexity analogues of other classical complexity classes have many applications, as surveyed in \cite{goos18landscape}. A grand challenge is to separate the communication complexity polynomial hierarchy. A notable subclass of the polynomial hierarchy is $\MA$, corresponding to Merlin--Arthur communication: Merlin (who knows both Alice's and Bob's inputs) sends a message (witness) to Alice and Bob (jointly constituting Arthur), who then run a randomized communication protocol with each other. On a $1$-input, some witness makes Alice and Bob accept with probability $\ge 2/3$. On a $0$-input, every witness makes Alice and Bob accept with probability $\le 1/3$. The classic problems \cp{Disjointness} and \cp{Inner Product Mod $2$} have Merlin--Arthur communication complexity $\widetilde{O}(\sqrt{N})$ \cite{aaronson09algebrization,chen20hardness} and $\Omega(\sqrt{N})$ \cite{klauck03rectangle}. No $\omega(\sqrt{N})$ lower bound is known for the Merlin--Arthur communication complexity of any explicit function. Also, $\MA$ is contained in $\AM$ (Arthur--Merlin) and in $\SP{2}$ (symmetric alternation), which are subclasses of the communication polynomial hierarchy for which no $\omega(\log N)$ lower bound is known for any explicit function. Thus $\MA$ is close to the frontier of communication complexity lower bounds. Also, Merlin--Arthur communication has applications to streaming \cite{chakrabarti14annotations,gur15arthur,ghosh24new}, property testing \cite{gur18non}, hardness of approximation \cite{abboud17distributed}, and cryptography \cite{bhadauria25snark}.

$\NP^\BPP$ is the class of all $F\colon\{0,1\}^N\times\{0,1\}^N\to\{0,1,\bot\}$ with $\NP$-type protocols that can make adaptive oracle queries to some $G\colon\{0,1\}^M\times\{0,1\}^M\to\{0,1,\bot\}$ in $\BPP$ with $M\in\poly N$. For every valid input to $F$ and every possible witness, the queries must be valid inputs to $G$. A randomized protocol could handle the whole sequence of queries, so there might as well be just one query, consisting of the input and the witness: $\NP^\BPP$ equals the class $\N\bigcdot\BPP$ (also denoted $\exists\bigcdot\BPP$) of all $F\colon\{0,1\}^N\times\{0,1\}^N\to\{0,1,\bot\}$ such that for some $G\colon\{0,1\}^{N+W}\times\{0,1\}^{N+W}\to\{0,1,\bot\}$ in $\BPP$ with $W\in\polylog N$, and for all valid inputs $(x,y)$ to $F$, \[\textstyle F(x,y)=1~\Rightarrow~\exists w:~G(xw,yw)=1\hspace*{1cm}F(x,y)=0~\Rightarrow~\forall w:~G(xw,yw)=0\] and $(xw,yw)$ is a valid input to $G$ for all $w$. (For surveys of complexity class dot operators, see \cite{zachos03combinatory} and \cite[lecture G]{kozen06theory}.) The $\NP^\BPP$ requirement that queries are valid corresponds to the $\N\bigcdot\BPP$ requirement that $(xw,yw)$ is valid. At a glance, $\N\bigcdot\BPP$ looks like $\MA$ where Merlin sends $w$ and Alice and Bob run the randomized protocol for $G(xw,yw)$. When $F(x,y)=1$, some $w$ makes Alice and Bob accept with probability $\ge 2/3$, but for each other $w$, their acceptance probability must be either $\ge 2/3$ or $\le 1/3$ (it doesn't matter which) for $\N\bigcdot\BPP$ but may be anything for $\MA$. This is the connection to pseudodeterminism. We have $\NP^\BPP=\N\bigcdot\BPP\subseteq\MA$. We answer one of the remaining open questions mentioned in \cite{goos18landscape}:

\begin{theorem} \label{thm:ma-np-bpp}
$\MA\not\subseteq\NP^\BPP$ in communication complexity.
\end{theorem}

Consider time complexity classes for a moment: Under a standard derandomization assumption, $\MA$ collapses to $\NP$ and hence equals $\NP^\BPP$. But the distinction still matters since some results such as $\EXP\subseteq\Ppoly~\Rightarrow~\EXP=\MA$ \cite{babai91nondeterministic} aren't known with $\NP^\BPP$ in place of $\MA$. \cite{fenner03oracle} proved the relativized separation $\MA\not\subseteq\NP^\BPP$. In general, relativized separations of time complexity classes are implied by separations of analogous query complexity classes, but the converse doesn't necessarily hold. In this case, the proof in \cite{fenner03oracle} doesn't yield $\MA\not\subseteq\NP^\BPP$ in query complexity, but the latter separation was shown by Mika G\"o\"os and reported in \cite{watson20quadratic}.

The upshot of \autoref{thm:ma-np-bpp} is that we contribute to the goal of better understanding the communication polynomial hierarchy, by picking apart its subclasses.

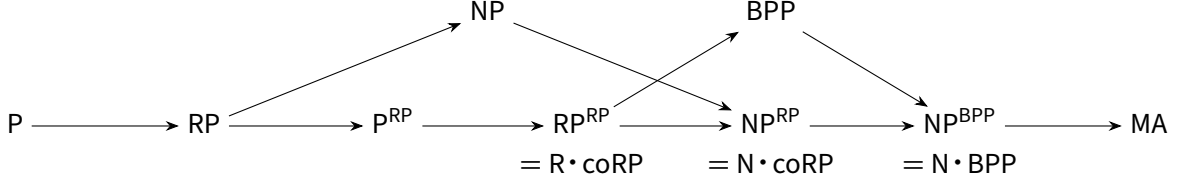
\begin{figure}[t]
\centering
\def\myw{2.5}
\def\myh{1.5}
\begin{tikzpicture}[every edge/.append style=mydir]
\node[myclassmap] (p) at (-1*\myw,0*\myh) {\strut $\P$};
\node[myclassmap] (rp) at (0*\myw,0*\myh) {\strut $\RP$};
\node[myclassmap] (prp) at (1*\myw,0*\myh) {\strut $\P^\RP$};
\node[myclassmap] (rprp) at (2*\myw,0*\myh) {\strut $\RP^\RP$};
\node[myclassmap] (rcorp) at (2*\myw,-0.35*\myh) {\strut $=\R\bigcdot\coRP$};
\node[myclassmap] (np) at (1.5*\myw,1*\myh) {\strut $\NP$};
\node[myclassmap] (bpp) at (3*\myw,1*\myh) {\strut $\BPP$};
\node[myclassmap] (nprp) at (3*\myw,0*\myh) {\strut $\NP^\RP$};
\node[myclassmap] (ncorp) at (3*\myw,-0.35*\myh) {\strut $=\N\bigcdot\coRP$};
\node[myclassmap] (npbpp) at (4*\myw,0*\myh) {\strut $\NP^\BPP$};
\node[myclassmap] (nbpp) at (4*\myw,-0.35*\myh) {\strut $=\N\bigcdot\BPP$};
\node[myclassmap] (ma) at (5*\myw,0*\myh) {\strut $\MA$};
\draw
(p) edge (rp)
(rp) edge (prp)
(rp) edge (np)
(prp) edge (rprp)
(rprp) edge (bpp)
(rprp) edge (nprp)
(np) edge (nprp)
(bpp) edge (npbpp)
(nprp) edge (npbpp)
(npbpp) edge (ma);
\end{tikzpicture}
\caption[Inclusions among communication complexity classes, where $\mathcal{C}_1\longrightarrow\mathcal{C}_2$ means $\mathcal{C}_1\subseteq\mathcal{C}_2$]{Inclusions among communication complexity classes, where $\mathcal{C}_1$\!\!\begin{tikzpicture}\node (l) at (0,0) {}; \node (r) at (1,0) {}; \draw (l) edge[-{Stealth[length=5pt]}] (r);\end{tikzpicture}\!\!$\mathcal{C}_2$ means $\mathcal{C}_1\subseteq\mathcal{C}_2$}
\label{fig:classes}
\end{figure}

\autoref{fig:classes}
provides some context regarding the theme of randomized oracles in communication complexity. We'd like to highlight the open problem of proving $\BPP\not\subseteq\NP^\RP$. The weaker separation $\BPP\not\subseteq\NP^{\cp{Equality}}$ is known \cite{pitassi23strength}, but neither of the two weaker separations $\BPP\not\subseteq\RP^\RP$ and $\NP^\BPP\not\subseteq\NP^\RP$ is known. Analogous to $\NP^\NP=\N\bigcdot\coNP=\SigmaP{2}$, we have the folklore facts $\RP^\RP=\R\bigcdot\coRP$ and $\NP^\RP=\N\bigcdot\coRP$.

$\BPP\subseteq\R\bigcdot\coRP$ would hold by essentially the proof of $\BPP\subseteq\SigmaP{2}$ in \cite{lautemann83bpp} if we removed the implicit one-sided-error pseudodeterminism in the definition of $\R\bigcdot\coRP$. \cite{muchnik96general,buhrman99one} proved the relativized separation of the time complexity classes $\BPP\not\subseteq\RP^\RP$, but the proof doesn't yield $\BPP\not\subseteq\RP^\RP$ in query complexity. Proving the latter separation would be a first step toward proving $\BPP\not\subseteq\RP^\RP$ in communication complexity.

Similarly, $\N\bigcdot\BPP\subseteq\N\bigcdot\coRP$ would hold if we removed the pseudodeterminism. This is just the observation that for $\MA$, two-sided error can be made one-sided (perfect completeness). By the way, $\N\bigcdot\coRP$ is also interesting since it contains the complement of \cp{List Non-Equality}, which is a classic total function exhibiting a quadratic separation between zero-sided-error randomized and deterministic communication complexities \cite[\S\hspace*{-0.2em}~4.1]{kushilevitz97communication}.

The class $\P^\RP$ is also notable. It contains \cp{Greater Than}. We have $\BPP\not\subseteq\P^\RP$ since $\BPP\not\subseteq\P^\NP$ \cite{papakonstantinou14overlays}. Restricting $\P^\RP$ to constant numbers of queries yields the randomized boolean hierarchy, which is known to be strict \cite{pitassi21nondeterministic}.


\subsection{Definitions} \label{sec:intro:def}

For basic background on communication complexity, see the textbooks \cite{kushilevitz97communication,jukna12boolean,rao20communication,watson26complexity}. We define communication models for computing two-party partial functions $F\colon\{0,1\}^N\times\{0,1\}^N\to\{0,1,\bot\}$ where $\bot$ means ``don't care,'' and if $F(x,y)\ne\bot$ we say $(x,y)$ is a valid input. For each model, the corresponding complexity measure of $F$ is the minimum cost of any protocol computing $F$ in the model, and the corresponding complexity class is the set of all infinite families of $F$s with complexity $\polylog N$. For example, $\P$ is the class of all families with deterministic communication complexity $\polylog N$. If $D$ is a distribution over deterministic protocols $P$ (that output $1$ or $0$), we define the notation $D(x,y)=\Pr_{P\sim D}\bigbrackets{P(x,y)=1}$.

An $\RP$-type (one-sided-error randomized) protocol of cost $d$ for $F$ is a distribution $D$ over deterministic protocols of depth $d$ such that for all valid $(x,y)$: \[F(x,y)=1~\Rightarrow~D(x,y)\ge 2/3\hspace*{1cm}F(x,y)=0~\Rightarrow~D(x,y)=0\]

A $\BPP$-type (two-sided-error randomized) protocol of cost $d$ for $F$ is a distribution $D$ over deterministic protocols of depth $d$ such that for all valid $(x,y)$: \[F(x,y)=1~\Rightarrow~D(x,y)\ge 2/3\hspace*{1cm}F(x,y)=0~\Rightarrow~D(x,y)\le 1/3\]

A $\psP$-type (two-sided-error pseudodeterministic) protocol of cost $d$ for $F$ is a distribution $D$ over deterministic protocols of depth $d$ such that for all valid or invalid $(x,y)$, we have $D(x,y)\in[0,1/3]\cup[2/3,1]$, and: \[F(x,y)=1~\Rightarrow~D(x,y)\ge 2/3\hspace*{1cm}F(x,y)=0~\Rightarrow~D(x,y)\le 1/3\]

An $\NP$-type (nondeterministic) protocol of cost $d$ for $F$ is a set $\calP$ of $\le 2^d$ deterministic protocols of depth $d$ such that for all valid $(x,y)$: \[F(x,y)=1~\Rightarrow~\exists P\in\calP:~P(x,y)=1\hspace*{1cm}F(x,y)=0~\Rightarrow~\forall P\in\calP:~P(x,y)=0\]

An $\MA$-type (Merlin--Arthur) protocol of cost $d$ for $F$ is a set $\calD$ of $\le 2^d$ distributions over deterministic protocols of depth $d$ such that for all valid $(x,y)$: \[F(x,y)=1~\Rightarrow~\exists D\in\calD:~D(x,y)\ge 2/3\hspace*{1cm}F(x,y)=0~\Rightarrow~\forall D\in\calD:~D(x,y)\le 1/3\]

An $\N\bigcdot\BPP$-type protocol of cost $d$ for $F$ is a set $\calD$ of $\le 2^d$ distributions over deterministic protocols of depth $d$ such that for all valid $(x,y)$, we have $D(x,y)\in[0,1/3]\cup[2/3,1]$ for all $D\in\calD$, and: \[F(x,y)=1~\Rightarrow~\exists D\in\calD:~D(x,y)\ge 2/3\hspace*{1cm}F(x,y)=0~\Rightarrow~\forall D\in\calD:~D(x,y)\le 1/3\]

An $\N\bigcdot\coRP$-type protocol of cost $d$ for $F$ is a set $\calD$ of $\le 2^d$ distributions over deterministic protocols of depth $d$ such that for all valid $(x,y)$, we have $D(x,y)\in[0,1/3]\cup\{1\}$ for all $D\in\calD$, and: \[F(x,y)=1~\Rightarrow~\exists D\in\calD:~D(x,y)=1\hspace*{1cm}F(x,y)=0~\Rightarrow~\forall D\in\calD:~D(x,y)\le 1/3\]

An $\R\bigcdot\coRP$-type protocol of cost $d$ for $F$ is a distribution $D_1$ over distributions $D_2$ over deterministic protocols of depth $d$ such that for all valid $(x,y)$, we have $D_2(x,y)\in[0,1/3]\cup\{1\}$ for all $D_2$ in $D_1$'s support, and: \[F(x,y)=1~\Rightarrow~\Pr_{D_2\sim D_1}\bigbrackets{D_2(x,y)=1}\ge 2/3\hspace*{1cm}F(x,y)=0~\Rightarrow~\Pr_{D_2\sim D_1}\bigbrackets{D_2(x,y)\le 1/3}=1\]

The error probabilities in these models can be amplified as usual. In particular, amplifying from $1/3$ to any constant $>0$ only affects the cost by a constant factor.


\section{Approach} \label{sec:approach}


\subsection{Intuition} \label{sec:approach:intuition}

\emph{Query-to-communication lifting} is a general approach to deducing communication complexity lower bounds from query complexity (decision tree depth) lower bounds. The idea is to compose an outer function $f\colon\{0,1\}^n\to\{0,1,\bot\}$ with a two-party gadget function $g\colon\{0,1\}^h\times\{0,1\}^h\to\{0,1\}$ to define the two-party function $F\colon(\{0,1\}^h)^n\times(\{0,1\}^h)^n\to\{0,1,\bot\}$ by $F(x,y)=f\bigparens{g(x_1,y_1)\cdots g(x_n,y_n)}$. A lifting theorem says $F$'s communication complexity is at least $f$'s query complexity if $g$ is complicated enough to effectively preclude communication protocols from doing anything more clever than just running a decision tree for $f$ and straightforwardly evaluating $g(x_i,y_i)$ when the decision tree queries its $i$th input bit. The query lower bound for $f$ is the ``spherical cow in a vacuum'' version of the communication lower bound for $F$. The lifting theorem handles all the messiness of arbitrary communication protocols.

We have lifting theorems for many models of query/communication complexity, such as deterministic ($\P$), nondeterministic ($\NP$), randomized ($\BPP$), and so on \cite{raz99separation,goos16rectangles,goos15lower,goos18deterministic,goos19query,goos20query,garg20monotone,watson20zpp,chattopadhyay21query,pitassi21nondeterministic,lovett22lifting}. In these settings, any old query lower bound proof is fine, and the lifting theorem is used as a black box. For other models, we don't know a general lifting theorem, but we may still be able to apply lifting techniques in a white-box manner to imitate a particular query lower bound proof for a particular $f\fs$. Examples of white-box query-to-communication lifting include \cite{watson16nonnegative,yang24communication,huang25min,goos26pseudodeterministic}.

We don't have lifting theorems for $\psP$ or $\NP^\BPP$. If we did, our theorems would automatically follow from the corresponding query complexity separations, which are simple to prove. So we use the white-box approach, like \cite{goos26pseudodeterministic}. But the simplest proofs of $\ZPP\not\subseteq\psP$ and $\MA\not\subseteq\NP^\BPP$ in query complexity aren't ``liftable.'' For example: The proof of $\MA\not\subseteq\NP^\BPP$ in query complexity in \cite{watson20quadratic} reaches a contradiction of some randomized decision tree accepting some input $z$ with probability $\approx 1/2$ when it should be $\ge 2/3$ or $\le 1/3$. In the composed function $F$, $z$ corresponds to a \emph{slice} of inputs $\bigbracescolon{(x,y)}{g(x_1,y_1)\cdots g(x_n,y_n)=z}$. The acceptance probabilities might be $\ge 2/3$ for half of the slice's inputs and $\le 1/3$ for the other half, leading to average acceptance probability $\approx 1/2$ over the slice, without yielding a contradiction. So we first adapt the query complexity lower bound proofs to make them amenable to lifting.

Definitions: A randomized decision tree of cost $d$ is a distribution $D$ over deterministic decision trees $T$ of depth $d$. For an input $z\in\{0,1\}^n$, define the notation $D(z)=\Pr_{T\sim D}\bigbrackets{T(z)=1}$. Let $|z|$ be $z$'s weight (number of $1$s), and let $z_\leftarrow=z_1\cdots z_{n/2}$ and $z_\rightarrow=z_{n/2+1}\cdots z_n$ be $z$'s left and right halves.

\paragraph{Intuition for $\text{\sffamily\upshape\fontseries{sb}\selectfont ZPP}\not\subseteq\text{\sffamily\upshape\fontseries{sb}\selectfont psP}$:} To prove the query complexity version, define $f\colon\{0,1\}^n\to\{0,1,\bot\}$ by: \[f(z)~=~\begin{cases}1&\text{if $|z_\rightarrow|\ge n/3$~\,and~\,$|z_\leftarrow|=0$}\\0&\text{if $|z_\leftarrow|\ge n/3$~\,and~\,$|z_\rightarrow|=0$}\end{cases}\]

$f\in\ZPP$: We have $f\in\RP$ by the randomized decision tree that picks a uniformly random $i\in\{n/2+1,\ldots,n\}$ and queries and outputs $z_i$. We have $\overline{f}\in\RP$ by the randomized decision tree that picks a uniformly random $i\in\{1,\ldots,n/2\}$ and queries and outputs $z_i$.

$f\not\in\psP$: Suppose for contradiction that some total function $f'\colon\{0,1\}^n\to\{0,1\}$ that agrees with $f$ has a $1/3$-error randomized decision tree $D$ of cost $n/20$. Initially let $z=0^n$, and assume $f'(z)=1$ so $D(z)\ge 2/3$. A symmetric argument handles the case $f'(z)=0$. If we obtain $z'$ from $z$ by flipping a uniformly random $0$ to $1$ in $z_\leftarrow$, then $\Pr_{z',\,T\sim D}\bigbrackets{\text{$T(z')$ queries the flipped bit}}\le (n/20)/(n/2)<1/3$, so $\Ex_{z'}\bigbrackets{D(z')}=\Pr_{z',\,T\sim D}\bigbrackets{T(z')=1}\ge\Pr_{z',\,T\sim D}\bigbrackets{\text{$T(z)=1$ and $T(z')$ doesn't query the flipped bit}}>D(z)-1/3\ge 1/3$. Hence there exists an outcome $z'$ such that $D(z')>1/3$, so $f'(z')=1$ and thus $D(z')\ge 2/3$. Update $z$ to $z'$. Repeat this process, maintaining the invariant $f'(z)=1$, until $n/3$ many $0$s have been flipped to $1$s in $z_\leftarrow$. Then $f(z)=0$ and thus $f'(z)=0$, which is a contradiction.

This iterative argument alternates between ``sneak another $1$ into $z_\leftarrow$ without decreasing $D(z)$ much'' and ``boost $D(z)$ back up to $\ge 2/3$.'' The latter happens automatically as long as $D(z)>1/3$. But it's not automatic when we imitate this argument in communication complexity: We maintain the invariant that $F'(x,y)=1$ for a large fraction of all $(x,y)$ in $z$'s slice. We can sneak another $1$ into $z_\leftarrow$ without decreasing this fraction much, but the decreases would accumulate, so we need to interleave boosting steps. A key idea from \cite{goos26pseudodeterministic} and its progenitor \cite{gavinsky25unambiguous} exploits the hypothesized pseudodeterministic protocol to boost the fraction of $1$-inputs of $F'$ in $z$'s slice by restricting to some subrectangle. This restriction fixes (hardwires) some of $z$'s bits, so we can't subsequently flip any fixed $0$s to $1$s. This threatens our goal of getting $|z_\leftarrow|\ge n/3$. The next idea from \cite{gavinsky25unambiguous,goos26pseudodeterministic} is to instead randomly sneak $\sqrt{n}$ many $1$s at a time into $z_\leftarrow$ to offset the fixed $0$s, so $z_\leftarrow$ still ends up with many more $1$s than $0$s. If the pseudodeterministic protocol has cost $0.01\sqrt{n}$ (a weaker but good enough lower bound), then typically only $\approx 0.01\sqrt{n}$ bits of $z$ get fixed per iteration, in which case the protocol ``hardly notices'' the $\sqrt{n}$ many bits being flipped (\`a la birthday paradox). Some iterations may fix many more than $0.01\sqrt{n}$ bits, but that can't happen too often since it causes a corresponding decrease in a certain potential function.

Our proof is similar in outline to \cite{goos26pseudodeterministic} but simpler in the technical details. We iteratively whittle down a set of inputs to the outer function $f$ like in \cite{goos26pseudodeterministic}, but a fundamental difference is that we also focus on an individual input $z$ in this set, iteratively changing which input we're focusing on. To accomplish this, we employ the query-to-communication lifting tool from \cite{goos16rectangles} rather than the one in \cite{goos26pseudodeterministic}.

\paragraph{Intuition for $\text{\sffamily\upshape\fontseries{sb}\selectfont MA}\not\subseteq\text{\sffamily\upshape\fontseries{sb}\selectfont NP}^\text{\sffamily\upshape\fontseries{sb}\selectfont BPP}$:} To prove the query complexity version (in a less straightforward but more ``liftable'' way than in \cite{watson20quadratic}): Viewing the input $z$ as a square binary matrix, first define $f$ by $f(z)=1$ if some row is all-$1$s, and $f(z)=0$ if every row is at most a third $1$s. This $f$ is canonically in $\MA$, where rows correspond to Merlin's witnesses and columns correspond to Arthur's randomness. One row will play a distinguished role, and each other row will be all-$1$s or all-$0$s and thus might as well be a single bit. So redefine $f\colon\{0,1\}^n\to\{0,1,\bot\}$ by letting $z_\leftarrow$ be the distinguished row and $z_\rightarrow$ be all the ``single bit rows'': \[f(z)~=~\begin{cases}1&\text{if $|z_\leftarrow|=n/2$~\,\mywidthboxc{and}{or}~\,$|z_\rightarrow|>0$}\\0&\text{if $|z_\leftarrow|\le n/6$~\,and~\,$|z_\rightarrow|=0$}\end{cases}\]

$f\not\in\NP^\BPP=\N\bigcdot\BPP$: Suppose for contradiction $f$ has an $\N\bigcdot\BPP$-type decision tree of cost $\delta n$ where the $\BPP$ part has been amplified to have error probability $\varepsilon$ instead of $1/3$, where $\delta=\varepsilon=1/20$. Initially let $z=1^{n/2}0^{n/2}$, so $f(z)=1$. Fixing the pertinent witness for input $z$ yields a randomized decision tree $D$ of cost $\delta n$ that has acceptance probability $\ge 1-\varepsilon$ for input $z$, and $\le\varepsilon$ for each $0$-input of $f\fs$, and $\ge 1-\varepsilon$ or $\le\varepsilon$ for each other $1$-input of $f\fs$. Similar to $\ZPP\not\subseteq\psP$, we iteratively modify the initial $1$-input $z$ to reach some $0$-input by sneaking a total of $n/3$ many $0$s into $z_\leftarrow$ (and leaving $z_\rightarrow$ alone), maintaining the invariant that $D(z)\ge 2/3$. At the end, $f(z)=0$ and thus $D(z)\le\varepsilon$, contradicting the invariant. In each iteration, randomly sneaking another $0$ into $z_\leftarrow$ only decreases $D(z)$ by $\le\delta n/(n/6)=6\delta$, so $D(z)\ge 2/3-6\delta>1/3$ after this. We claim that $D(z)$ is automatically boosted back to $\ge 2/3$, but this is trickier to see because this intermediate $z$ is an invalid input to $f$ and thus $D(z)$ is unconstrained.

Proof of the claim: Consider the set $Z$ of all ``nearby'' $1$-inputs $z'$ obtained from the current $z$ (after flipping a $1$ to $0$ in $z_\leftarrow$) by flipping a $0$ to $1$ in $z_\rightarrow$. We have $\Pr_{z'\in Z,\,T\sim D}\bigbrackets{T(z)\ne T(z')}\le\Pr_{z'\in Z,\,T\sim D}\bigbrackets{\text{$T(z')$ queries the flipped bit}}\le\delta n/|Z|=2\delta$ and thus $D(z)\in\Ex_{z'\in Z}\bigbrackets{D(z')}\pm 2\delta$. Since $D$ behaves pseudodeterministically on $Z$, it's an $\varepsilon$-error randomized decision tree for some $f'\colon Z\to\{0,1\}$. We have $\Ex_{z'\in Z}\bigbrackets{D(z')}\in\Ex_{z'\in Z}\bigbrackets{f'(z')}\pm\varepsilon$. There exists a deterministic decision tree $T$ in the support of $D$ such that $\Pr_{z'\in Z}\bigbrackets{f'(z')\ne T(z')}\le\varepsilon$ and thus $\Ex_{z'\in Z}\bigbrackets{f'(z')}\in\Ex_{z'\in Z}\bigbrackets{T(z')}\pm\varepsilon$. We have $\Pr_{z'\in Z}\bigbrackets{T(z')\ne T(z)}\le\Pr_{z'\in Z}\bigbrackets{\text{$T(z')$ queries the flipped bit}}\le\delta n/|Z|=2\delta$ and thus $\Ex_{z'\in Z}\bigbrackets{T(z')}\in T(z)\pm 2\delta$. Stringing everything together, $D(z)\in T(z)\pm(4\delta+2\varepsilon)\subseteq T(z)\pm 1/3$. If $T(z)=0$ then $D(z)\le 1/3$ contradicting $D(z)>1/3$, so we must have $T(z)=1$ and thus $D(z)\ge 2/3$. In summary, $D(z)>1/3$ is automatically boosted to $D(z)\ge 2/3$, and we continue to the next iteration.

We imitate this proof in communication complexity using query-to-communication lifting techniques, like with $\ZPP\not\subseteq\psP$. We sneak $\sqrt{n}$ many $0$s into $z_\leftarrow$ per iteration, to offset the fixed $1$s, of which there are typically only $\approx 0.01\sqrt{n}$ per iteration, assuming the protocol has cost $0.01\sqrt{n}$. The most subtle aspect of the proof is in the analogue to the boosting step: This requires restricting to a subrectangle that fixes some bits of the input to $f\fs$. Doing this naively might result in a fixed $1$ in the right half of the input (preventing us from reaching a $0$-input at the end of the proof) since the boosting focuses on the inputs $z'\in Z$, which each have a $1$ in their right half. Intuitively, the restriction must focus on $z$ itself to maintain $z_\rightarrow=0^{n/2}$, but must accomplish boosting on $Z$. This involves a nuanced back-and-forth between $z$ and $Z$, like how the query complexity proof considers $T(z)$ even though $T$ was chosen to work on $Z$.


\subsection{The lifting lemma} \label{sec:approach:lifting}

We use the technical centerpiece of \cite{goos16rectangles}. To state it, we need some definitions. Assume $h=100\log n$ and $m\le n$. Let $S=\{0,1\}^h\times\{0,1\}^h$ and $S^m=(\{0,1\}^h)^m\times(\{0,1\}^h)^m$. Define $g\colon S\to\{0,1\}$ by $g(x,y)=(x\cdot y)\bmod 2$, and define $g^m\colon S^m\to\{0,1\}^m$ by $g^m(x,y)=\bigparens{g(x_1,y_1),\ldots,g(x_m,y_m)}$. For any rectangle $R\subseteq S^m$ and $z\in\{0,1\}^m$, define $R_z=\bigbracescolon{(x,y)\in R}{g^m(x,y)=z}$. A rectangle $R\subseteq S^m$ is \emph{balanced} iff for each $z$, $|R_z|$ is within a factor $1.01$ of $|R|/2^m$. A basic fact is that $S^m$ itself is balanced, since $|S_1|=(1-2^{-h})|S|/2$ and so for every $z$, $|S^m_\sus{z}|=|S_\sus{z_1}|\cdots|S_\sus{z_m}|$ and $|S^m|/2^m=(|S|/2)^m$ are within a factor $1/(1-2^{-h})^m\le 1/(1-m2^{-h})\le 1/(1-n2^{-100\log n})=1/(1-n^{-99})\le 1.01$.

A \emph{structure} of \emph{width $w$} is a triple of partial assignments $(a,b,c)\in(\{0,1,*\}^h)^m\times(\{0,1,*\}^h)^m\times\{0,1,*\}^m$ (where $*$ means unassigned) such that for $w$ many $i\in[m]$, $(a_i,b_i,c_i)$ is totally assigned and $g(a_i,b_i)=c_i$, and for all other $i\in[m]$, $(a_i,b_i,c_i)$ is totally unassigned. Let $S^{a,b}\subseteq S^m$ be the rectangle of all $(x,y)$ that agree with $(a,b)$, and note that $g^m(S^{a,b})\subseteq\{0,1\}^m$ is the set of all $z$ that agree with $c$. A rectangle $R\subseteq S^m$ is \emph{$(a,b,c)$-balanced} iff $R\subseteq S^{a,b}$ and for each $z$ that agrees with $c$, $|R_z|$ is within a factor $1.01$ of $|R|/2^{m-w}$.

\begin{lemma}[\cite{goos16rectangles}] \label{lem:packing}
For every rectangle $R\subseteq S^m$ and $z\in\{0,1\}^m$ with $|R_z|\ge 2^{-k}|S^m_\sus{z}|$, there exists a set $\calQ$ of pairwise disjoint subrectangles of $R$ such that $\sum_{Q\in\calQ}|Q_z|\ge 0.99|R_z|$ and for each $Q\in\calQ$ there exists a structure $(a,b,c)$ of width $w\le k+8$ such that $z$ agrees with $c$, $Q$ is $(a,b,c)$-balanced, and $|Q_z|\ge 2^{-(k+8-w)}|S^{a,b}_\sus{z}|$.
\end{lemma}

(Note to readers familiar with the ``packing algorithm'' in \cite{goos16rectangles}: A tiny tweak is needed to get \autoref{lem:packing} as stated. The proof in \cite{goos16rectangles} yields $|Q_z|\ge 2^{-(k-10w\log n)}|S^{a,b}_\sus{z}|$ if $w>0$ (and we drop the insignificant factor $10\log n$ for the sake of tidiness) and $|Q_z|\ge 2^{-(k+8+\log n)}|S^m_\sus{z}|$ if $w=0$, which can only happen for one $Q$, namely the remaining rectangle at the end of the top level of the packing algorithm. To get rid of the $+\log n$ for that pesky $Q$, we just halt the top level partitioning and discard the remaining $Q$ when $|Q_z|<2^{-8}|R_z|$. This only contributes $2^{-8}<0.005$ to the algorithm's error parameter, which we set as $\varepsilon=0.01$.)

More notation: For any $z\in\{0,1\}^m$ and set of indices $I\subseteq[m]$, let $z^I$ be $z$ but with the bits indexed by $I$ flipped. For any rectangle $R\subseteq S^m$ and set $\calI$ of sets of indices, let $R_\sus{z^\calI}=\bigcup_{I\in\calI}R_\sus{z^I}$.
\begin{figure}[t]
\centering
\begin{tikzpicture}
\draw[thick] (0,0) rectangle (6,6);
\draw[thick,fill=mylight] (1.5,6) -- (0.5,0) -- (1.25,0) -- (2.25,6) -- cycle;
\draw[thick,fill=mylight] (3.25,6) -- (2.25,0) -- (3,0) -- (4,6) -- cycle;
\draw[thick,fill=mylight] (4,6) -- (3,0) -- (3.75,0) -- (4.75,6) -- cycle;
\draw[thick,fill=mylight] (4.75,6) -- (3.75,0) -- (4.5,0) -- (5.5,6) -- cycle;
\draw[thick,fill=mygray] (0.5,1.5) rectangle (5.5,4.5);
\draw[thick,fill=mydark] (1.25,4.5) -- (0.75,1.5) -- (1.5,1.5) -- (2,4.5) -- cycle;
\draw[thick,fill=mydark] (3,4.5) -- (2.5,1.5) -- (3.25,1.5) -- (3.75,4.5) -- cycle;
\draw[thick,fill=mydark] (3.75,4.5) -- (3.25,1.5) -- (4,1.5) -- (4.5,4.5) -- cycle;
\draw[thick,fill=mydark] (4.5,4.5) -- (4,1.5) -- (4.75,1.5) -- (5.25,4.5) -- cycle;
\node at (0.35,5.7) {$S^m_\sus{\phantom{z}}$};
\node at (1.825,5.7) {$S^m_\sus{z}$};
\node at (4.325,5.7) {$S^m_{\smash{\raisebox{-1.5pt}{\scriptsize $z^\calI$}}}$};
\node at (0.85,4.2) {$R_{\phantom{z}}$};
\node at (1.6,4.2) {$R_z$};
\node at (4.1,4.2) {$R_\sus{z^\calI}$};
\draw (3.82,4.25) edge[-{Stealth[length=5pt]}] (2.97,4.25);
\draw (4.38,4.25) edge[-{Stealth[length=5pt]}] (5.19,4.25);
\draw (4.06,5.75) edge[-{Stealth[length=5pt]}] (3.22,5.75);
\draw (4.59,5.75) edge[-{Stealth[length=5pt]}] (5.44,5.75);
\end{tikzpicture}
\caption{Illustration of $R_z$ and $R_\sus{z^\calI}$}
\label{fig:slices}
\end{figure}
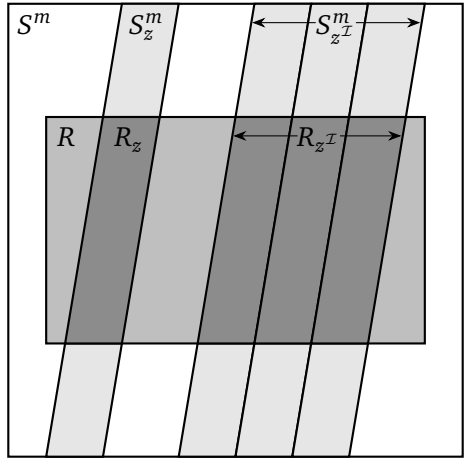
See \autoref{fig:slices}.


\section{\texorpdfstring{$\text{\sffamily\upshape\fontseries{sb}\selectfont ZPP}\not\subseteq\text{\sffamily\upshape\fontseries{sb}\selectfont psP}$}{ZPP is not in psP}} \label{sec:zpp-psp}

We prove \autoref{thm:zpp-psp}. For $z\in\{0,1\}^n$, let $|z|$ be $z$'s weight (number of $1$s), and let $z_\leftarrow=z_1\cdots z_{n/2}$ and $z_\rightarrow=z_{n/2+1}\cdots z_n$ be $z$'s left and right halves. Define $f\colon\{0,1\}^n\to\{0,1,\bot\}$ by: \[f(z)~=~\begin{cases}1&\text{if $|z_\rightarrow|\ge n/3$~\,and~\,$|z_\leftarrow|=0$}\\0&\text{if $|z_\leftarrow|\ge n/3$~\,and~\,$|z_\rightarrow|=0$}\end{cases}\] With the setup from \autoref{sec:approach:lifting}, define $F\colon S^n\to\{0,1,\bot\}$ by $F(x,y)=f(g^n(x,y))$.

$F\in\ZPP$: We have $F\in\RP$ by the randomized protocol that picks a uniformly random $i\in\{n/2+1,\ldots,n\}$ and computes and outputs $g(x_i,y_i)$. We have $\overline{F}\in\RP$ by the randomized protocol that picks a uniformly random $i\in\{1,\ldots,n/2\}$ and computes and outputs $g(x_i,y_i)$.

$F\not\in\psP$: Suppose for contradiction that some total function $F'\colon S^n\to\{0,1\}$ that agrees with $F$ has a randomized protocol $D$ of cost $d=0.01\sqrt{n}$ that's been amplified to have two-sided error probability $0.01$ instead of $1/3$. Let $F_1=F^{-1}(1)$ and $F_0=F^{-1}(0)$ and $F'_\sus{1}=F^{\prime-1}(1)$ and $F'_\sus{0}=F^{\prime-1}(0)$. Let $z=0^n$ and assume $|F'_\sus{1}\cap S^n_\sus{z}|\ge 0.5|S^n_\sus{z}|$. A symmetric argument handles the case $|F'_\sus{0}\cap S^n_\sus{z}|\ge 0.5|S^n_\sus{z}|$.

We describe a process that iteratively updates $z\in\{0,1\}^n$, a structure $(a,b,c)$ of width $w$, a rectangle $R\subseteq S^n$, and $k\ge 0$, maintaining these invariants: \[|z_\rightarrow|=0\hspace*{0.6cm}\text{$z$ agrees with $c$}\hspace*{0.6cm}\text{$R$ is $(a,b,c)$-balanced}\hspace*{0.6cm}|R_z|\ge 2^{-k}|S^{a,b}_\sus{z}|\hspace*{0.6cm}|F'_\sus{1}\cap R_z|\ge 0.5|R_z|\] Initially, $z=0^n$, $(a,b,c)$ is all $*$s, $w=0$, $R=S^n$, and $k=0$. Each iteration extends $(a,b,c)$ and shrinks $R$ to a subrectangle. Each iteration is either \emph{safe}, in which case $\sqrt{n}$ many $0$s flip to $1$s in $z_\leftarrow$ and $k$ increases by $\le d+13$, or \emph{unsafe}, in which case $z$ doesn't change and $k$ decreases by $\ge(\text{increase in $w$})/2>0$. This continues until $|z_\leftarrow|\ge n/3$, in other words, $(n/3)/\sqrt{n}=\sqrt{n}/3$ safe iterations. Thus at the end, we have $f(z)=0$ and therefore $R_z\subseteq F_0\subseteq F'_\sus{0}$, contradicting the invariant $|F'_\sus{1}\cap R_z|\ge 0.5|R_z|>0$. Since the safe iterations increase $k$ by $\le(d+13)\sqrt{n}/3$ in total, and the unsafe iterations decrease $k$ by $\ge w/2$ in total, and since $k\ge 0$, we always have $w/2\le(d+13)\sqrt{n}/3$, which implies $w\le 0.02n$ since $d=0.01\sqrt{n}$.

We now describe an iteration. Since $c_i\ne *$ for $\le w\le 0.02n$ many $i\le n/2$, and $z_i=1$ for $\le n/3+\sqrt{n}$ many $i\le n/2$, we have $c_i=*$ and $z_i=0$ for $\ge n/2-0.02n-(n/3+\sqrt{n})\ge 0.1n$ many $i\le n/2$. Let $\calI$ be the set of all sets $I\subseteq\bracescolon{i\le n/2}{\text{$c_i=*$ and $z_i=0$}}$ with $|I|=\sqrt{n}$.

Since $D$ is a $0.01$-error randomized protocol for $F'$, there exists a deterministic protocol $P$ in the support of $D$ such that $P(x,y)\ne F'(x,y)$ with probability $\le 0.01$ over $(x,y)$ sampled from the uniform mixture of the uniform distribution over $R_z$ and the uniform distribution over $R_\sus{z^\calI}$. Hence $P(x,y)\ne F'(x,y)$ for $\le 0.02|R_z|$ many $(x,y)\in R_z$ and for $\le 0.02|R_\sus{z^\calI}|$ many $(x,y)\in R_\sus{z^\calI}$. Let $P_1=P^{-1}(1)$ and $P_0=P^{-1}(0)$. Since $P$ has depth $d$, $P_1\cap R$ is partitioned into at most $2^d$ rectangles $R^1,R^2,R^3,\ldots{}$. Say $\ell$ is \emph{major} if $|R^\ell_\sus{z}|\ge 2^{-(d+4)}|R_z|$ and \emph{minor} otherwise. For each major $\ell$, we have $|R^\ell_\sus{z}|\ge 2^{-(k+d+4)}|S^{a,b}_\sus{z}|$, so by \autoref{lem:packing} (applied to the projections of $R^\ell$ and $z$ to the $*$ coordinates, with $m=n-w$), there exists a set $\calQ^\ell$ of pairwise disjoint subrectangles of $R^\ell$ such that $\sum_\sus{Q\in\calQ^\ell}|Q_z|\ge 0.99|R^\ell_\sus{z}|$ and for each $Q\in\calQ^\ell$ there exists a structure $(a',b',c')$ of width $w'$ extending $(a,b,c)$ such that $z$ agrees with $c'$, $Q$ is $(a',b',c')$-balanced, and $|Q_z|\ge 2^{-((k+d+4)+8-(w'-w))}|S^{a',b'}_\sus{z}|$. We have
\begin{align*}
\textstyle|F'_\sus{1}\cap P_0\cap R_z|~&\textstyle\le~0.02|R_z|\\
\textstyle\sum_{\text{minor }\ell}|R^\ell_\sus{z}|~&\textstyle\le~2^d\cdot 2^{-(d+4)}|R_z|~\le~0.07|R_z|\\
\textstyle\sum_{\text{major }\ell}\bigparens{|R^\ell_\sus{z}|-\sum_\sus{Q\in\calQ^\ell}|Q_z|}~&\textstyle\le~\sum_{\text{major }\ell}0.01|R^\ell_\sus{z}|~\le~0.01|R_z|
\end{align*}
and thus, letting $\calQ=\bigcup_{\text{major }\ell}\calQ^\ell$:
\begin{align*}
\textstyle\sum_{Q\in\calQ}|Q_z|~&\textstyle=~|(F'_\sus{1}\cup P_1)\cap R_z|-|F'_\sus{1}\cap P_0\cap R_z|-\sum_{\text{minor }\ell}|R^\ell_\sus{z}|-\sum_{\text{major }\ell}\bigparens{|R^\ell_\sus{z}|-\sum_\sus{Q\in\calQ^\ell}|Q_z|}\\
&\textstyle\ge~0.5|R_z|-0.02|R_z|-0.07|R_z|-0.01|R_z|\\
&\textstyle=~0.4|R_z|
\end{align*}
Say $Q\in\calQ$ is \emph{good} if $w'-w\le 3d$ (for the $w'$ associated with $Q$) and \emph{bad} otherwise. This iteration is \emph{safe} if $\sum_{\text{good }Q\in\calQ}|Q_z|\ge 0.2|R_z|$ and \emph{unsafe} otherwise, in which case $\sum_{\text{bad }Q\in\calQ}|Q_z|\ge 0.2|R_z|$.

\paragraph{Unsafe iteration:} We have \[\textstyle\sum_{\text{bad }Q\in\calQ}|F'_\sus{0}\cap Q_z|~\le~|F'_\sus{0}\cap P_1\cap R_z|~\le~0.02|R_z|~\le~0.1\sum_{\text{bad }Q\in\calQ}|Q_z|\] so there exists a bad $Q\in\calQ$ with associated $(a',b',c')$ and $w'$ such that $|F'_\sus{0}\cap Q_z|\le 0.1|Q_z|\le 0.5|Q_z|$, so $|F'_\sus{1}\cap Q_z|\ge 0.5|Q_z|$. We conclude the iteration by updating $(a,b,c)\leftarrow(a',b',c')$ and $w\leftarrow w'$ and $R\leftarrow Q$ and $k\leftarrow k+d+12-(w'-w)$. The invariants are maintained. As we wanted for an unsafe iteration, $z$ doesn't change and $k$ decreases by $(\text{increase in $w$})-d-12\ge(\text{increase in $w$})/2$ since $(\text{increase in $w$})>3d$ since $Q$ is bad.

\paragraph{Safe iteration:} For each good $Q\in\calQ$ with associated $(a',b',c')$ and $w'$, if we sample a uniformly random $I\in\calI$ and a uniformly random permutation of $I$'s elements, then{\allowdisplaybreaks
\begin{align*}
\textstyle\Pr[\text{$z^I$ doesn't agree with $c'$}]~&\textstyle=~\Pr\bigbrackets{(\exists i\in I):~c'_\sus{i}=0}\\*
&\textstyle=~\Pr\bigbrackets{(\exists i\,:\,\text{$c_i=*$ and $c'_\sus{i}=0$})~(\exists e\in[\sqrt{n}]):~\text{$I$'s $e$th element is $i$}}\\
&\textstyle\le~\sum_\sus{i\,:\,c_i=*\text{ and }c'_\sus{i}=0}\sum_\sus{e\in[\sqrt{n}]}\Pr[\text{$I$'s $e$th element is $i$}]\\
&\textstyle\le~(w'-w)\cdot\sqrt{n}\cdot 1/0.1n\\
&\textstyle\le~3d\cdot 10/\sqrt{n}\\*
&\textstyle=~0.3
\end{align*}
}since $I$'s $e$th element is uniformly distributed over a set of $\ge 0.1n$ many $i$, and since $w'-w\le 3d$ since $Q$ is good. If $z^I$ agrees with $c'$, then $|Q_\sus{z^I}|\ge 0.99|Q|/2^{n-w'}\ge 0.98|Q_z|$ since $z$ agrees with $c'$ and $Q$ is $(a',b',c')$-balanced. Thus: \[\textstyle|Q_\sus{z^\calI}|~=~\sum_{I\in\calI}|Q_\sus{z^I}|~\ge~\sum_\sus{I\in\calI\,:\,z^I\text{ agrees with }c'}|Q_\sus{z^I}|~\ge~0.7|\calI|\cdot 0.98|Q_z|~\ge~0.6|\calI|\cdot|Q_z|\] It follows that: \[\textstyle\sum_{\text{good }Q\in\calQ}|Q_\sus{z^\calI}|~\ge~0.6|\calI|\sum_{\text{good }Q\in\calQ}|Q_z|~\ge~0.6|\calI|\cdot 0.2|R_z|~=~0.12|\calI|\cdot|R_z|\] We have $|R_z|\ge 0.99|R|/2^{n-w}\ge 0.98|R_\sus{z^I}|$ for each $I\in\calI$ since $z$ and $z^I$ agree with $c$ and $R$ is $(a,b,c)$-balanced, so: \[\textstyle|R_z|~\ge~0.98\sum_{I\in\calI}|R_\sus{z^I}|/|\calI|~=~0.98|R_\sus{z^\calI}|/|\calI|\] It follows that: \[\textstyle\sum_{\text{good }Q\in\calQ}|Q_\sus{z^\calI}|~\ge~0.12|\calI|\cdot|R_z|~\ge~0.12\cdot 0.98|R_\sus{z^\calI}|~\ge~0.1|R_\sus{z^\calI}|\] We have \[\textstyle\sum_{\text{good }Q\in\calQ}|F'_\sus{0}\cap Q_\sus{z^\calI}|~\le~|F'_\sus{0}\cap P_1\cap R_\sus{z^\calI}|~\le~0.02|R_\sus{z^\calI}|~\le~0.2\sum_{\text{good }Q\in\calQ}|Q_\sus{z^\calI}|\] so there exists a good $Q\in\calQ$ with associated $(a',b',c')$ and $w'$ such that $|F'_\sus{0}\cap Q_\sus{z^\calI}|\le 0.2|Q_\sus{z^\calI}|\le 0.5|Q_\sus{z^\calI}|$, so $|F'_\sus{1}\cap Q_\sus{z^\calI}|\ge 0.5|Q_\sus{z^\calI}|$. Hence there exists an $I\in\calI$ such that $|F'_\sus{1}\cap Q_\sus{z^I}|\ge 0.5|Q_\sus{z^I}|$ and $z^I$ agrees with $c'$ (since $Q_\sus{z^I}=\emptyset$ if $z^I$ doesn't agree with $c'$). Also: \[\textstyle|Q_\sus{z^I}|~\ge~0.98|Q_z|~\ge~0.98\cdot 2^{-(k+d+12)}|S^{a',b'}_\sus{z}|~\ge~0.98\cdot 2^{-(k+d+12)}\cdot 0.98|S^{a',b'}_\sus{\raisebox{-1.5pt}{\scriptsize $z^I$}}|~\ge~2^{-(k+d+13)}|S^{a',b'}_\sus{\raisebox{-1.5pt}{\scriptsize $z^I$}}|\] We conclude the iteration by updating $z\leftarrow z^I$ and $(a,b,c)\leftarrow (a',b',c')$ and $w\leftarrow w'$ and $R\leftarrow Q$ and $k\leftarrow k+d+13$. The invariants are maintained. As we wanted for a safe iteration, $\sqrt{n}$ many $0$s flip to $1$s in $z_\leftarrow$ and $k$ increases by $\le d+13$.


\section{\texorpdfstring{$\text{\sffamily\upshape\fontseries{sb}\selectfont MA}\not\subseteq\text{\sffamily\upshape\fontseries{sb}\selectfont NP}^\text{\sffamily\upshape\fontseries{sb}\selectfont BPP}$}{MA is not in NP\textasciicircum BPP}} \label{sec:ma-np-bpp}

We prove \autoref{thm:ma-np-bpp}. For $z\in\{0,1\}^n$, let $|z|$ be $z$'s weight (number of $1$s), and let $z_\leftarrow=z_1\cdots z_{n/2}$ and $z_\rightarrow=z_{n/2+1}\cdots z_n$ be $z$'s left and right halves. Define $f\colon\{0,1\}^n\to\{0,1,\bot\}$ by: \[f(z)~=~\begin{cases}1&\text{if $|z_\leftarrow|=n/2$~\,\mywidthboxc{and}{or}~\,$|z_\rightarrow|>0$}\\0&\text{if $|z_\leftarrow|\le n/6$~\,and~\,$|z_\rightarrow|=0$}\end{cases}\] With the setup from \autoref{sec:approach:lifting}, define $F\colon S^n\to\{0,1,\bot\}$ by $F(x,y)=f(g^n(x,y))$.

$F\in\MA$: Consider the randomized protocol that picks a uniformly random $i\in\{1,\ldots,n/2\}$ and computes and outputs $g(x_i,y_i)$, and for each $i\in\{n/2+1,\ldots,n\}$ consider the deterministic protocol that computes and outputs $g(x_i,y_i)$. This set of $n/2+1$ protocols is an $\MA$-type protocol of cost $O(\log n)$ for $F$: If $F(x,y)=1$ then at least one of these protocols outputs $1$ with probability $1\ge 2/3$, and if $F(x,y)=0$ then each of these protocols outputs $1$ with probability $\le 1/3$.

$F\not\in\NP^\BPP=\N\bigcdot\BPP$: Suppose for contradiction $F$ has an $\N\bigcdot\BPP$-type protocol $\calD$ of cost $d=0.01\sqrt{n}$ where the $\BPP$ part has been amplified to have error probability $0.01$ instead of $1/3$. The proof has a \emph{preprocessing phase} and an \emph{iterative phase}. For $z_\leftarrow$, we use the notation $i$, $I$, $\calI$ for indices, sets of indices, and sets of sets of indices. For $z_\rightarrow$, we use the notation $j$, $J$, $\calJ$ for indices, sets of indices, and sets of sets of indices.

\paragraph{Preprocessing phase:} For $z=1^{n/2}0^{n/2}$ we have $f(z)=1$, so $F(x,y)=1$ for all $(x,y)\in S^n_\sus{z}$. Since $|\calD|\le 2^d$, there exists a $D\in\calD$ such that $D(x,y)\ge 0.99$ for $\ge 2^{-d}|S^n_\sus{z}|$ many $(x,y)\in S^n_\sus{z}$. Define $F'$ to have the same valid inputs as $F$, and such that $D$ is a $0.01$-error randomized protocol for $F'$. Let $F_1=F^{-1}(1)$ and $F_0=F^{-1}(0)$ and $F'_\sus{1}=F^{\prime-1}(1)$ and $F'_\sus{0}=F^{\prime-1}(0)$. Thus $|F'_\sus{1}\cap S^n_\sus{z}|\ge 2^{-d}|S^n_\sus{z}|$ and $F_1\subseteq F'_\sus{1}\cup F'_\sus{0}$ and $F_0\subseteq F'_\sus{0}$.

Let $D'$ be the further amplified version of $D$ that outputs the majority vote of $2d$ independent runs of $D$, so $D'$ is a $0.01\cdot 2^{-d}$-error randomized protocol for $F'$ of cost $2d^2$. There exists a deterministic protocol $P$ in the support of $D'$ such that $P(x,y)\ne F'(x,y)$ with probability $\le 0.01\cdot 2^{-d}$ over $(x,y)$ sampled from the uniform mixture of the uniform distribution over $S^n_\sus{z}$ and the uniform distribution over $S^n_\sus{z^\calJ}$ where $\calJ=\bigbraces{\{n/2+1\},\ldots,\{n\}}$, noting that $S^n_\sus{z}\cup S^n_\sus{z^\calJ}\subseteq F_1\subseteq F'_\sus{1}\cup F'_\sus{0}$. Hence $P(x,y)\ne F'(x,y)$ for $\le 0.02\cdot 2^{-d}|S^n_\sus{z}|$ many $(x,y)\in S^n_\sus{z}$ and for $\le 0.02\cdot 2^{-d}|S^n_\sus{z^\calJ}|$ many $(x,y)\in S^n_\sus{z^\calJ}$. Let $P_1=P^{-1}(1)$ and $P_0=P^{-1}(0)$. Since $P$ has depth $2d^2$, $P_1$ is partitioned into at most $2^{2d^2}$ rectangles $R^1,R^2,R^3,\ldots{}$. Say $\ell$ is \emph{major} if $|R^\ell_\sus{z}|\ge 2^{-(d+2d^2+4)}|S^n_\sus{z}|$ and \emph{minor} otherwise. For each major $\ell$, by \autoref{lem:packing} (with $m=n$), there exists a set $\calQ^\ell$ of pairwise disjoint subrectangles of $R^\ell$ such that $\sum_\sus{Q\in\calQ^\ell}|Q_z|\ge 0.99|R^\ell_\sus{z}|$ and for each $Q\in\calQ^\ell$ there exists a structure $(a,b,c)$ of width $w\le d+2d^2+12\le 0.01n$ such that $z$ agrees with $c$, $Q$ is $(a,b,c)$-balanced, and $|Q_z|\ge 2^{-(d+2d^2+12-w)}|S^{a,b}_\sus{z}|$. We have
\begin{align*}
\textstyle|F'_\sus{1}\cap P_0\cap S^n_\sus{z}|~&\textstyle\le~0.02\cdot 2^{-d}|S^n_\sus{z}|~\le~0.02|F'_\sus{1}\cap S^n_\sus{z}|\\
\textstyle\sum_{\text{minor }\ell}|R^\ell_\sus{z}|~&\textstyle\le~2^{2d^2}\cdot 2^{-(d+2d^2+4)}|S^n_\sus{z}|~\le~0.07\cdot 2^{-d}|S^n_\sus{z}|~\le~0.07|F'_\sus{1}\cap S^n_\sus{z}|\\
\textstyle\sum_{\text{major }\ell}\bigparens{|R^\ell_\sus{z}|-\sum_\sus{Q\in\calQ^\ell}|Q_z|}~&\textstyle\le~\sum_{\text{major }\ell}0.01|R^\ell_\sus{z}|~\le~0.01|P_1\cap S^n_\sus{z}|
\end{align*}
and thus, letting $\calQ=\bigcup_{\text{major }\ell}\calQ^\ell$:
\begin{align*}
\textstyle\sum_{Q\in\calQ}|Q_z|~&\textstyle=~|(F'_\sus{1}\cup P_1)\cap S^n_\sus{z}|-|F'_\sus{1}\cap P_0\cap S^n_\sus{z}|-\sum_{\text{minor }\ell}|R^\ell_\sus{z}|-\sum_{\text{major }\ell}\bigparens{|R^\ell_\sus{z}|-\sum_\sus{Q\in\calQ^\ell}|Q_z|}\\
&\textstyle\ge~|(F'_\sus{1}\cup P_1)\cap S^n_\sus{z}|(1-0.02-0.07-0.01)\\
&\textstyle\ge~0.9|F'_\sus{1}\cap S^n_\sus{z}|\\
&\textstyle\ge~0.9\cdot 2^{-d}|S^n_\sus{z}|
\end{align*}
For each $Q\in\calQ$ with associated $(a,b,c)$ and $w$, and for $J=\{j\}\in\calJ$, $z^J$ agrees with $c$ iff $c_j=*$, which holds for $\ge n/2-w\ge 0.98|\calJ|$ many $J\in\calJ$. If $z^J$ agrees with $c$, then $|Q_\sus{z^J}|\ge 0.99|Q|/2^{n-w}\ge 0.98|Q_z|$ since $z$ agrees with $c$ and $Q$ is $(a,b,c)$-balanced. Thus: \[\textstyle|Q_\sus{z^\calJ}|~=~\sum_{J\in\calJ}|Q_\sus{z^J}|~\ge~\sum_\sus{J\in\calJ\,:\,z^J\text{ agrees with }c}|Q_\sus{z^J}|~\ge~0.98|\calJ|\cdot 0.98|Q_z|~\ge~0.9|\calJ|\cdot|Q_z|\] It follows that: \[\textstyle\sum_{Q\in\calQ}|Q_\sus{z^\calJ}|~\ge~0.9|\calJ|\sum_{Q\in\calQ}|Q_z|~\ge~0.9|\calJ|\cdot 0.9\cdot 2^{-d}|S^n_\sus{z}|~\ge~0.8|\calJ|\cdot 2^{-d}|S^n_\sus{z}|\] We have $|S^n_\sus{z}|\ge 0.99|S^n|/2^n\ge 0.98|S^n_\sus{z^J}|$ for each $J\in\calJ$ since $S^n$ is balanced, so: \[\textstyle|S^n_\sus{z}|~\ge~0.98\sum_{J\in\calJ}|S^n_\sus{z^J}|/|\calJ|~=~0.98|S^n_\sus{z^\calJ}|/|\calJ|\] It follows that: \[\textstyle\sum_{Q\in\calQ}|Q_\sus{z^\calJ}|~\ge~0.8|\calJ|\cdot 2^{-d}|S^n_\sus{z}|~\ge~0.8\cdot 2^{-d}\cdot 0.98|S^n_\sus{z^\calJ}|~\ge~0.7\cdot 2^{-d}|S^n_\sus{z^\calJ}|\] We have \[\textstyle\sum_{Q\in\calQ}|F'_\sus{0}\cap Q_\sus{z^\calJ}|~\le~|F'_\sus{0}\cap P_1\cap S^n_\sus{z^\calJ}|~\le~0.02\cdot 2^{-d}|S^n_\sus{z^\calJ}|~\le~0.03\sum_{Q\in\calQ}|Q_\sus{z^\calJ}|\] so there exists a $Q\in\calQ$ such that $|F'_\sus{0}\cap Q_\sus{z^\calJ}|\le 0.03|Q_\sus{z^\calJ}|\le 0.3|Q_\sus{z^\calJ}|$, so $|F'_\sus{1}\cap Q_\sus{z^\calJ}|\ge 0.7|Q_\sus{z^\calJ}|$. Define $R$ to be this $Q$, with associated $(a,b,c)$ and $w$, and redefine $\calJ=\bigbracescolon{\{j\}}{\text{$j>n/2$ and $c_j=*$}}$, which doesn't change $R_\sus{z^\calJ}$ since $R$ is $(a,b,c)$-balanced. In summary, letting $k=3d^2-w$: \[\textstyle z=1^{n/2}0^{n/2}\hspace*{0.6cm}\text{$z$ agrees with $c$}\hspace*{0.6cm}\text{$R$ is $(a,b,c)$-balanced}\hspace*{0.6cm}|R_z|\ge 2^{-k}|S^{a,b}_\sus{z}|\hspace*{0.6cm}|F'_\sus{1}\cap R_\sus{z^\calJ}|\ge 0.7|R_\sus{z^\calJ}|\]

\paragraph{Iterative phase:} We describe a process that iteratively updates $z\in\{0,1\}^n$, a structure $(a,b,c)$ of width $w$, a rectangle $R\subseteq S^n$, and $k\ge 0$, maintaining these invariants, where $\calJ=\bigbracescolon{\{j\}}{\text{$j>n/2$ and $c_j=*$}}$: \[|z_\rightarrow|=0\hspace*{0.6cm}\text{$z$ agrees with $c$}\hspace*{0.6cm}\text{$R$ is $(a,b,c)$-balanced}\hspace*{0.6cm}|R_z|\ge 2^{-k}|S^{a,b}_\sus{z}|\hspace*{0.6cm}|F'_\sus{1}\cap R_\sus{z^\calJ}|\ge 0.7|R_\sus{z^\calJ}|\] The initial $z=1^{n/2}0^{n/2}$, $(a,b,c)$, $w$, $R$, and $k=3d^2-w$ are from the end of the preprocessing phase. Each iteration extends $(a,b,c)$ and shrinks $R$ to a subrectangle. Each iteration is either \emph{safe}, in which case $\sqrt{n}$ many $1$s flip to $0$s in $z_\leftarrow$ and $k$ increases by $\le d+13$, or \emph{unsafe}, in which case $z$ doesn't change and $k$ decreases by $\ge(\text{increase in $w$})/2>0$. This continues until $|z_\leftarrow|\le n/6$, in other words $(n/3)/\sqrt{n}=\sqrt{n}/3$ safe iterations. Thus at the end, we have $f(z)=0$ and therefore $R_z\subseteq F_0\subseteq F'_\sus{0}$, which leads to a contradiction with $|F'_\sus{1}\cap R_\sus{z^\calJ}|\ge 0.7|R_\sus{z^\calJ}|$, as we show at the end of the proof. Since the preprocessing and safe iterations increase $k$ by $\le 3d^2+(d+13)\sqrt{n}/3$ in total, and the preprocessing and unsafe iterations decrease $k$ by $\ge w/2$ in total, and since $k\ge 0$, we always have $w/2\le 3d^2+(d+13)\sqrt{n}/3$, which implies $w\le 0.02n$ since $d=0.01\sqrt{n}$.

A key issue is that to maintain the invariant $|F'_\sus{1}\cap R_\sus{z^\calJ}|\ge 0.7|R_\sus{z^\calJ}|$, we must consider the behavior of subrectangles with respect to $z^\calJ$, but to maintain the invariant $|z_\rightarrow|=0$, we must apply \autoref{lem:packing} with respect to $z$. Reconciling this involves a subtle back-and-forth between $R_z$ and $R_\sus{z^\calJ}$.

We now describe an iteration. Since $c_i\ne *$ for $\le w\le 0.02n$ many $i\le n/2$, and $z_i=0$ for $\le n/3$ many $i\le n/2$, we have $c_i=*$ and $z_i=1$ for $\ge n/2-0.02n-n/3\ge 0.1n$ many $i\le n/2$. Let $\calI$ be the set of all sets $I\subseteq\bracescolon{i\le n/2}{\text{$c_i=*$ and $z_i=1$}}$ with $|I|=\sqrt{n}$. We have $c_j\ne *$ for $\le w\le 0.02n$ many $j>n/2$, and $\calJ$ is the set of all sets $J\subseteq\bracescolon{j>n/2}{c_j=*}$ with $|J|=1$. Let $\calI\times\calJ=\bracescolon{I\cup J}{\text{$I\in\calI$ and $J\in\calJ$}}$. Note that $R_z\cup R_\sus{z^\calI}$ generally consists of invalid inputs (on which we know nothing a priori about $D$'s behavior), but $R_\sus{z^\calJ}\cup R_\sus{z^{\calI\times\calJ}}\subseteq F_1\subseteq F'_\sus{1}\cup F'_\sus{0}$.

Since $D$ is a $0.01$-error randomized protocol for $F'$, there exists a deterministic protocol $P$ in the support of $D$ such that $P(x,y)\ne F'(x,y)$ with probability $\le 0.01$ over $(x,y)$ sampled from the uniform mixture of the uniform distribution over $R_\sus{z^\calJ}$ and the uniform distribution over $R_\sus{z^{\calI\times\calJ}}$. Hence $P(x,y)\ne F'(x,y)$ for $\le 0.02|R_\sus{z^\calJ}|$ many $(x,y)\in R_\sus{z^\calJ}$ and for $\le 0.02|R_\sus{z^{\calI\times\calJ}}|$ many $(x,y)\in R_\sus{z^{\calI\times\calJ}}$. Let $P_1=P^{-1}(1)$ and $P_0=P^{-1}(0)$. Since $P$ has depth $d$, $P_1\cap R$ is partitioned into at most $2^d$ rectangles $R^1,R^2,R^3,\ldots{}$. Say $\ell$ is \emph{major} if $|R^\ell_\sus{z}|\ge 2^{-(d+4)}|R_z|$ and \emph{minor} otherwise. For each major $\ell$, we have $|R^\ell_\sus{z}|\ge 2^{-(k+d+4)}|S^{a,b}_\sus{z}|$, so by \autoref{lem:packing} (applied to the projections of $R^\ell$ and $z$ to the $*$ coordinates, with $m=n-w$), there exists a set $\calQ^\ell$ of pairwise disjoint subrectangles of $R^\ell$ such that $\sum_\sus{Q\in\calQ^\ell}|Q_z|\ge 0.99|R^\ell_\sus{z}|$ and for each $Q\in\calQ^\ell$ there exists a structure $(a',b',c')$ of width $w'$ extending $(a,b,c)$ such that $z$ agrees with $c'$, $Q$ is $(a',b',c')$-balanced, and $|Q_z|\ge 2^{-((k+d+4)+8-(w'-w))}|S^{a',b'}_\sus{z}|$. The latter bound implies $w'\le k+d+12+w\le 0.04n$. We have
\begin{align*}
\textstyle\sum_{\text{minor }\ell}|R^\ell_\sus{z}|~&\textstyle\le~2^d\cdot 2^{-(d+4)}|R_z|~\le~0.07|R_z|\\
\textstyle\sum_{\text{major }\ell}\bigparens{|R^\ell_\sus{z}|-\sum_\sus{Q\in\calQ^\ell}|Q_z|}~&\textstyle\le~\sum_{\text{major }\ell}0.01|R^\ell_\sus{z}|~\le~0.01|R_z|
\end{align*}
and thus, letting $\calQ=\bigcup_{\text{major }\ell}\calQ^\ell$:
\begin{equation} \label{eq:a}
\textstyle\sum_{Q\in\calQ}|Q_z|~=~|P_1\cap R_z|-\sum_{\text{minor }\ell}|R^\ell_\sus{z}|-\sum_{\text{major }\ell}\bigparens{|R^\ell_\sus{z}|-\sum_\sus{Q\in\calQ^\ell}|Q_z|}~\ge~|P_1\cap R_z|-0.08|R_z|\tag{$\star$}
\end{equation}
For each $Q\in\calQ$ with associated $(a',b',c')$ and $w'$, and for $J=\{j\}\in\calJ$, $z^J$ agrees with $c'$ iff $c'_\sus{j}=*$, which holds for $\ge n/2-w'\ge 0.92(n/2)\ge 0.92|\calJ|$ many $J\in\calJ$. If $z^J$ agrees with $c'$, then $|Q_\sus{z^J}|\ge 0.99|Q|/2^{n-w'}\ge 0.98|Q_z|$ since $z$ agrees with $c'$ and $Q$ is $(a',b',c')$-balanced. Thus:
\begin{equation} \label{eq:b}
\textstyle|Q_\sus{z^\calJ}|~=~\sum_{J\in\calJ}|Q_\sus{z^J}|~\ge~\sum_\sus{J\in\calJ\,:\,z^J\text{ agrees with }c'}|Q_\sus{z^J}|~\ge~0.92|\calJ|\cdot 0.98|Q_z|~\ge~0.9|\calJ|\cdot|Q_z|\tag{$\dagger$}
\end{equation}
It follows that: \[\textstyle|P_1\cap R_\sus{z^\calJ}|~\ge~\sum_{Q\in\calQ}|Q_\sus{z^\calJ}|~\ge~0.9|\calJ|\sum_{Q\in\calQ}|Q_z|~\ge~0.9|\calJ|\bigparens{|P_1\cap R_z|-0.08|R_z|}\] By the symmetry between outputting $1$ and outputting $0$, we also have: \[\textstyle|P_0\cap R_\sus{z^\calJ}|~\ge~0.9|\calJ|\bigparens{|P_0\cap R_z|-0.08|R_z|}\] We have $|R_z|\ge 0.99|R|/2^{n-w}\ge 0.98|R_\sus{z^J}|$ for each $J\in\calJ$ since $R$ is $(a,b,c)$-balanced, so:
\begin{equation} \label{eq:c}
\textstyle|R_z|~\ge~0.98\sum_{J\in\calJ}|R_\sus{z^J}|/|\calJ|~=~0.98|R_\sus{z^\calJ}|/|\calJ|\tag{$\ddagger$}
\end{equation}
It follows that \[\textstyle|P_0\cap R_\sus{z^\calJ}|~\ge~0.9|\calJ|\bigparens{|P_0\cap R_z|/|R_z|-0.08}|R_z|~\ge~0.8\bigparens{|P_0\cap R_z|/|R_z|-0.08}|R_\sus{z^\calJ}|\] and thus: \[\textstyle|F'_\sus{0}\cap R_\sus{z^\calJ}|~\ge~|P_0\cap R_\sus{z^\calJ}|-|F'_\sus{1}\cap P_0\cap R_\sus{z^\calJ}|~\ge~0.8\bigparens{|P_0\cap R_z|/|R_z|-0.08}|R_\sus{z^\calJ}|-0.02|R_\sus{z^\calJ}|\] The invariant $|F'_\sus{1}\cap R_\sus{z^\calJ}|\ge 0.7|R_\sus{z^\calJ}|$ implies $|F'_\sus{0}\cap R_\sus{z^\calJ}|\le 0.3|R_\sus{z^\calJ}|$, so combining yields: \[\textstyle 0.3~\ge~0.8\bigparens{|P_0\cap R_z|/|R_z|-0.08}-0.02~\ge~0.8|P_0\cap R_z|/|R_z|-0.1\] Rearranging then yields $|P_0\cap R_z|\le 0.5|R_z|$, so $|P_1\cap R_z|\ge 0.5|R_z|$. As we showed in \eqref{eq:a}, this implies \[\textstyle\sum_{Q\in\calQ}|Q_z|~\ge~|P_1\cap R_z|-0.08|R_z|~\ge~0.5|R_z|-0.08|R_z|~\ge~0.4|R_z|\] Say $Q\in\calQ$ is \emph{good} if $w'-w\le 3d$ (for the $w'$ associated with $Q$) and \emph{bad} otherwise. This iteration is \emph{safe} if $\sum_{\text{good }Q\in\calQ}|Q_z|\ge 0.2|R_z|$ and \emph{unsafe} otherwise, in which case $\sum_{\text{bad }Q\in\calQ}|Q_z|\ge 0.2|R_z|$.

\paragraph{Unsafe iteration:} As we showed in \eqref{eq:b} and \eqref{eq:c}: \[\textstyle\sum_{\text{bad }Q\in\calQ}|Q_\sus{z^\calJ}|~\ge~0.9|\calJ|\sum_{\text{bad }Q\in\calQ}|Q_z|~\ge~0.9|\calJ|\cdot 0.2|R_z|~\ge~0.9\cdot 0.2\cdot 0.98|R_\sus{z^\calJ}|~\ge~0.1|R_\sus{z^\calJ}|\] We have \[\textstyle\sum_{\text{bad }Q\in\calQ}|F'_\sus{0}\cap Q_\sus{z^\calJ}|~\le~|F'_\sus{0}\cap P_1\cap R_\sus{z^\calJ}|~\le~0.02|R_\sus{z^\calJ}|~\le~0.2\sum_{\text{bad }Q\in\calQ}|Q_\sus{z^\calJ}|\] so there exists a bad $Q\in\calQ$ with associated $(a',b',c')$ and $w'$ such that $|F'_\sus{0}\cap Q_\sus{z^\calJ}|\le 0.2|Q_\sus{z^\calJ}|\le 0.3|Q_\sus{z^\calJ}|$, so $|F'_\sus{1}\cap Q_\sus{z^\calJ}|\ge 0.7|Q_\sus{z^\calJ}|$. We conclude the iteration by updating $(a,b,c)\leftarrow(a',b',c')$ and $w\leftarrow w'$ and $R\leftarrow Q$ and $k\leftarrow k+d+12-(w'-w)$ and $\calJ\leftarrow\bigbracescolon{\{j\}}{\text{$j>n/2$ and $c'_\sus{j}=*$}}$. The invariants are maintained. As we wanted for an unsafe iteration, $z$ doesn't change and $k$ decreases by $(\text{increase in $w$})-d-12\ge(\text{increase in $w$})/2$ since $(\text{increase in $w$})>3d$ since $Q$ is bad.

\paragraph{Safe iteration:} For each good $Q\in\calQ$ with associated $(a',b',c')$ and $w'$, if we sample a uniformly random $I\in\calI$ and a uniformly random permutation of $I$'s elements, then{\allowdisplaybreaks
\begin{align*}
\textstyle\Pr[\text{$z^I$ doesn't agree with $c'$}]~&\textstyle=~\Pr\bigbrackets{(\exists i\in I):~c'_\sus{i}=1}\\*
&\textstyle=~\Pr\bigbrackets{(\exists i\,:\,\text{$c_i=*$ and $c'_\sus{i}=1$})~(\exists e\in[\sqrt{n}]):~\text{$I$'s $e$th element is $i$}}\\
&\textstyle\le~\sum_\sus{i\,:\,c_i=*\text{ and }c'_\sus{i}=1}\sum_\sus{e\in[\sqrt{n}]}\Pr[\text{$I$'s $e$th element is $i$}]\\
&\textstyle\le~(w'-w)\cdot\sqrt{n}\cdot 1/0.1n\\
&\textstyle\le~3d\cdot 10/\sqrt{n}\\*
&\textstyle=~0.3
\end{align*}
}since $I$'s $e$th element is uniformly distributed over a set of $\ge 0.1n$ many $i$, and since $w'-w\le 3d$ since $Q$ is good. If $z^I$ agrees with $c'$, then for $J=\{j\}\in\calJ$, $z^{I\cup\,J}$ agrees with $c'$ iff $c'_\sus{j}=*$, which holds for $\ge n/2-w'\ge 0.92(n/2)\ge 0.92|\calJ|$ many $J\in\calJ$. Thus $z^{I\cup\,J}$ agrees with $c'$ for $\ge 0.7|\calI|\cdot 0.92|\calJ|\ge 0.62|\calI\times\calJ|$ many $I\cup J\in\calI\times\calJ$. If $z^{I\cup\,J}$ agrees with $c'$, then $|Q_\sus{z^{I\cup\,J}}|\ge 0.99|Q|/2^{n-w'}\ge 0.98|Q_z|$ since $z$ agrees with $c'$ and $Q$ is $(a',b',c')$-balanced. Thus:
\begin{align*}
\textstyle|Q_\sus{z^{\calI\times\calJ}}|~=~\sum_{I\cup\,J\in\calI\times\calJ}|Q_\sus{z^{I\cup\,J}}|~\ge~\sum_\sus{I\cup\,J\in\calI\times\calJ\,:\,z^{I\cup\,J}\text{ agrees with }c'}|Q_\sus{z^{I\cup\,J}}|~&\textstyle\ge~0.62|\calI\times\calJ|\cdot 0.98|Q_z|\\
&\textstyle\ge~0.6|\calI\times\calJ|\cdot|Q_z|
\end{align*}
It follows that: \[\textstyle\sum_{\text{good }Q\in\calQ}|Q_\sus{z^{\calI\times\calJ}}|~\ge~0.6|\calI\times\calJ|\sum_{\text{good }Q\in\calQ}|Q_z|~\ge~0.6|\calI\times\calJ|\cdot 0.2|R_z|~=~0.12|\calI\times\calJ|\cdot|R_z|\] We have $|R_z|\ge 0.99|R|/2^{n-w}\ge 0.98|R_\sus{z^{I\cup\,J}}|$ for each $I\cup J\in\calI\times\calJ$ since $z$ and $z^{I\cup\,J}$ agree with $c$ and $R$ is $(a,b,c)$-balanced, so: \[\textstyle|R_z|~\ge~0.98\sum_{I\cup\,J\in\calI\times\calJ}|R_\sus{z^{I\cup\,J}}|/|\calI\times\calJ|~=~0.98|R_\sus{z^{\calI\times\calJ}}|/|\calI\times\calJ|\] It follows that: \[\textstyle\sum_{\text{good }Q\in\calQ}|Q_\sus{z^{\calI\times\calJ}}|~\ge~0.12|\calI\times\calJ|\cdot|R_z|~\ge~0.12\cdot 0.98|R_\sus{z^{\calI\times\calJ}}|~\ge~0.1|R_\sus{z^{\calI\times\calJ}}|\] We have \[\textstyle\sum_{\text{good }Q\in\calQ}|F'_\sus{0}\cap Q_\sus{z^{\calI\times\calJ}}|~\le~|F'_\sus{0}\cap P_1\cap R_\sus{z^{\calI\times\calJ}}|~\le~0.02|R_\sus{z^{\calI\times\calJ}}|~\le~0.2\sum_{\text{good }Q\in\calQ}|Q_\sus{z^{\calI\times\calJ}}|\] so there exists a good $Q\in\calQ$ with associated $(a',b',c')$ and $w'$ such that $|F'_\sus{0}\cap Q_\sus{z^{\calI\times\calJ}}|\le 0.2|Q_\sus{z^{\calI\times\calJ}}|\le 0.3|Q_\sus{z^{\calI\times\calJ}}|$, so $|F'_\sus{1}\cap Q_\sus{z^{\calI\times\calJ}}|\ge 0.7|Q_\sus{z^{\calI\times\calJ}}|$. Hence there exists an $I\in\calI$ such that $|F'_\sus{1}\cap Q_\sus{(z^I)^\calJ}|\ge 0.7|Q_\sus{(z^I)^\calJ}|$ and $z^I$ agrees with $c'$ (since $Q_\sus{(z^I)^\calJ}=\emptyset$ if $z^I$ doesn't agree with $c'$). Also: \[\textstyle|Q_\sus{z^I}|~\ge~0.98|Q_z|~\ge~0.98\cdot 2^{-(k+d+12)}|S^{a',b'}_\sus{z}|~\ge~0.98\cdot 2^{-(k+d+12)}\cdot 0.98|S^{a',b'}_\sus{\raisebox{-1.5pt}{\scriptsize $z^I$}}|~\ge~2^{-(k+d+13)}|S^{a',b'}_\sus{\raisebox{-1.5pt}{\scriptsize $z^I$}}|\] We conclude the iteration by updating $z\leftarrow z^I$ and $(a,b,c)\leftarrow(a',b',c')$ and $w\leftarrow w'$ and $R\leftarrow Q$ and $k\leftarrow k+d+13$ and $\calJ\leftarrow\bigbracescolon{\{j\}}{\text{$j>n/2$ and $c'_\sus{j}=*$}}$. The invariants are maintained. As we wanted for a safe iteration, $\sqrt{n}$ many $1$s flip to $0$s in $z_\leftarrow$ and $k$ increases by $\le d+13$.

\paragraph{Contradiction:} After the last iteration, $f(z)=0$ and therefore $R_z\subseteq F_0\subseteq F'_\sus{0}$. Since $D$ is a $0.01$-error randomized protocol for $F'$, there exists a deterministic protocol $P$ in the support of $D$ such that $P(x,y)\ne F'(x,y)$ with probability $\le 0.01$ over $(x,y)$ sampled from the uniform mixture of the uniform distribution over $R_z$ and the uniform distribution over $R_\sus{z^\calJ}$. Hence $P(x,y)\ne F'(x,y)$ for $\le 0.02|R_z|$ many $(x,y)\in R_z$ and for $\le 0.02|R_\sus{z^\calJ}|$ many $(x,y)\in R_\sus{z^\calJ}$. Let $P_1=P^{-1}(1)$ and $P_0=P^{-1}(0)$. We argued in the iterative phase that the invariant $|F'_\sus{1}\cap R_\sus{z^\calJ}|\ge 0.7|R_\sus{z^\calJ}|$ and $|F'_\sus{1}\cap P_0\cap R_\sus{z^\calJ}|\le 0.02|R_\sus{z^\calJ}|$ imply $|P_1\cap R_z|\ge 0.5|R_z|$ in this case. This contradicts $|P_1\cap R_z|=|F'_\sus{0}\cap P_1\cap R_z|\le 0.02|R_z|$.


\section*{AI usage statement}

The author did not use AI for anything.

\small\bibliography{nbp}

@InProceedings{papakonstantinou14overlays,
  author    = {Periklis Papakonstantinou and Dominik Scheder and Hao Song},
  title     = {Overlays and Limited Memory Communication},
  booktitle = {Proceedings of the 29th Conference on Computational Complexity (CCC)},
  year      = {2014},
  pages     = {298--308},
  publisher = {IEEE},
  doi       = {10.1109/CCC.2014.37},
}

@Article{aaronson09algebrization,
  author  = {Scott Aaronson and Avi Wigderson},
  title   = {Algebrization: {A} New Barrier in Complexity Theory},
  journal = {ACM Transactions on Computation Theory},
  year    = {2009},
  volume  = {1},
  number  = {1},
  pages   = {2:1--2:54},
  doi     = {10.1145/1490270.1490272},
}

@Book{kushilevitz97communication,
  title     = {Communication Complexity},
  publisher = {Cambridge University Press},
  year      = {1997},
  author    = {Eyal Kushilevitz and Noam Nisan},
  doi       = {10.1017/CBO9780511574948},
}

@Book{rao20communication,
  title     = {Communication Complexity and Applications},
  publisher = {Cambridge University Press},
  year      = {2020},
  author    = {Anup Rao and Amir Yehudayoff},
  doi       = {10.1017/9781108671644},
}

@Book{jukna12boolean,
  title     = {Boolean Function Complexity: {A}dvances and Frontiers},
  publisher = {Springer},
  year      = {2012},
  author    = {Stasys Jukna},
  volume    = {27},
  series    = {Algorithms and Combinatorics},
  doi       = {10.1007/978-3-642-24508-4},
}

@Book{watson26complexity,
  title     = {Complexity in Computer Science},
  publisher = {Cambridge University Press},
  year      = {2026},
  author    = {Thomas Watson},
  doi       = {10.1017/9781009752329},
}

@Article{pitassi21nondeterministic,
  author  = {Toniann Pitassi and Morgan Shirley and Thomas Watson},
  title   = {Nondeterministic and Randomized Boolean Hierarchies in Communication Complexity},
  journal = {Computational Complexity},
  year    = {2021},
  volume  = {30},
  number  = {2},
  pages   = {10},
  doi     = {10.1007/S00037-021-00210-5},
}

@Article{lautemann83bpp,
  author  = {Clemens Lautemann},
  title   = {{BPP} and the Polynomial Hierarchy},
  journal = {Information Processing Letters},
  year    = {1983},
  volume  = {17},
  number  = {4},
  pages   = {215--217},
  doi     = {10.1016/0020-0190(83)90044-3},
}

@Article{watson20quadratic,
  author  = {Thomas Watson},
  title   = {Quadratic Simulations of {M}erlin--{A}rthur Games},
  journal = {ACM Transactions on Computation Theory},
  year    = {2020},
  volume  = {12},
  number  = {2},
  pages   = {14:1--14:11},
  doi     = {10.1145/3389399},
}

@Article{goos18landscape,
  author  = {Mika G{\"o}{\"o}s and Toniann Pitassi and Thomas Watson},
  title   = {The Landscape of Communication Complexity Classes},
  journal = {Computational Complexity},
  year    = {2018},
  volume  = {27},
  number  = {2},
  pages   = {245--304},
  doi     = {10.1007/s00037-018-0166-6},
}

@Article{goos16rectangles,
  author  = {Mika G{\"o}{\"o}s and Shachar Lovett and Raghu Meka and Thomas Watson and David Zuckerman},
  title   = {Rectangles Are Nonnegative Juntas},
  journal = {SIAM Journal on Computing},
  year    = {2016},
  volume  = {45},
  number  = {5},
  pages   = {1835--1869},
  doi     = {10.1137/15M103145X},
}

@InProceedings{goos26pseudodeterministic,
  author    = {Mika G{\"{o}}{\"{o}}s and Nathaniel Harms and Artur Riazanov and Anastasia Sofronova and Dmitry Sokolov and Weiqiang Yuan},
  title     = {Pseudodeterministic Communication Complexity},
  booktitle = {Proceedings of the 58th Symposium on Theory of Computing (STOC)},
  year      = {2026},
  pages     = {2030--2039},
  publisher = {ACM},
  doi       = {10.1145/3798129.3800907},
}

@Article{gavinsky25unambiguous,
  author  = {Dmytro Gavinsky},
  title   = {Unambiguous Parity-Query Complexity},
  journal = {Random Structures and Algorithms},
  year    = {2025},
  volume  = {66},
  number  = {3},
  doi     = {10.1002/RSA.70010},
}

@InProceedings{buhrman99one,
  author    = {Harry Buhrman and Lance Fortnow},
  title     = {One-Sided Versus Two-Sided Error in Probabilistic Computation},
  booktitle = {Proceedings of the 16th International Symposium on Theoretical Aspects of Computer Science (STACS)},
  year      = {1999},
  pages     = {100--109},
  publisher = {Springer},
  doi       = {10.1007/3-540-49116-3_9},
}

@Article{muchnik96general,
  author  = {Andrei Muchnik and Nikolai Vereshchagin},
  title   = {A General Method to Construct Oracles Realizing Given Relationships Between Complexity Classes},
  journal = {Theoretical Computer Science},
  year    = {1996},
  volume  = {157},
  number  = {2},
  pages   = {227--258},
  doi     = {10.1016/0304-3975(95)00161-1},
}

@Article{fenner03oracle,
  author  = {Stephen Fenner and Lance Fortnow and Stuart Kurtz and Lide Li},
  title   = {An Oracle Builder's Toolkit},
  journal = {Information and Computation},
  year    = {2003},
  volume  = {182},
  number  = {2},
  pages   = {95--136},
  doi     = {10.1016/S0890-5401(03)00018-X},
}

@Book{kozen06theory,
  title     = {Theory of Computation},
  publisher = {Springer},
  year      = {2006},
  author    = {Dexter Kozen},
  series    = {Texts in Computer Science},
  doi       = {10.1007/1-84628-477-5},
}

@InProceedings{zachos03combinatory,
  author    = {Stathis Zachos and Aris Pagourtzis},
  title     = {Combinatory Complexity: {O}perators on Complexity Classes},
  booktitle = {Proceedings of the 4th Panhellenic Logic Symposium (PLS)},
  year      = {2003},
}

@Article{goos19query,
  author  = {Mika G{\"o}{\"o}s and Pritish Kamath and Toniann Pitassi and Thomas Watson},
  title   = {Query-to-Communication Lifting for {$\textrm{P}^{\footnotesize\textrm{NP}}$}},
  journal = {Computational Complexity},
  year    = {2019},
  volume  = {28},
  number  = {1},
  pages   = {113--144},
  doi     = {10.1007/s00037-018-0175-5},
}

@Article{goos20query,
  author  = {Mika G{\"o}{\"o}s and Toniann Pitassi and Thomas Watson},
  title   = {Query-to-Communication Lifting for {BPP}},
  journal = {SIAM Journal on Computing},
  year    = {2020},
  volume  = {49},
  number  = {4},
  pages   = {FOCS17\-441--FOCS17\-461},
  doi     = {https://doi.org/10.1137/17M115339X},
}

@Article{goos18deterministic,
  author  = {Mika G{\"o}{\"o}s and Toniann Pitassi and Thomas Watson},
  title   = {Deterministic Communication vs.~Partition Number},
  journal = {SIAM Journal on Computing},
  year    = {2018},
  volume  = {47},
  number  = {6},
  pages   = {2435--2450},
  doi     = {10.1137/16M1059369},
}

@Article{watson20zpp,
  author  = {Thomas Watson},
  title   = {A {$\textrm{ZPP}^{\footnotesize\textrm{NP[1]}}$} Lifting Theorem},
  journal = {ACM Transactions on Computation Theory},
  year    = {2020},
  volume  = {12},
  number  = {4},
  pages   = {27:1--27:20},
  doi     = {10.1145/3428673},
}

@Article{garg20monotone,
  author  = {Ankit Garg and Mika G{\"{o}}{\"{o}}s and Pritish Kamath and Dmitry Sokolov},
  title   = {Monotone Circuit Lower Bounds from Resolution},
  journal = {Theory of Computing},
  year    = {2020},
  volume  = {16},
  pages   = {1--30},
  doi     = {10.4086/TOC.2020.V016A013},
}

@Article{chattopadhyay21query,
  author  = {Arkadev Chattopadhyay and Yuval Filmus and Sajin Koroth and Or Meir and Toniann Pitassi},
  title   = {Query-to-Communication Lifting Using Low-Discrepancy Gadgets},
  journal = {SIAM Journal on Computing},
  year    = {2021},
  volume  = {50},
  number  = {1},
  pages   = {171--210},
  doi     = {10.1137/19M1310153},
}

@TechReport{gat11probabilistic,
  author      = {Eran Gat and Shafi Goldwasser},
  title       = {Probabilistic Search Algorithms with Unique Answers and Their Cryptographic Applications},
  institution = {Electronic Colloquium on Computational Complexity (ECCC)},
  year        = {2011},
  number      = {TR11-136},
  url         = {https://eccc.weizmann.ac.il/report/2011/136/},
}

@InProceedings{huynh12virtue,
  author    = {Trinh Huynh and Jakob Nordstr{\"{o}}m},
  title     = {On the Virtue of Succinct Proofs: {A}mplifying Communication Complexity Hardness to Time-Space Trade-Offs in Proof Complexity},
  booktitle = {Proceedings of the 44th Symposium on Theory of Computing (STOC)},
  year      = {2012},
  pages     = {233--248},
  publisher = {ACM},
  doi       = {10.1145/2213977.2214000},
}

@Article{chen26polynomial,
  author  = {Lijie Chen and Zhenjian Lu and Igor Oliveira and Hanlin Ren and Rahul Santhanam},
  title   = {Polynomial-Time Pseudodeterministic Construction of Primes},
  journal = {Journal of the ACM},
  year    = {2026},
  volume  = {73},
  number  = {2},
  pages   = {14:1--14:43},
  doi     = {10.1145/3803408},
}

@InProceedings{lu21pseudodeterministic,
  author    = {Zhenjian Lu and Igor Oliveira and Rahul Santhanam},
  title     = {Pseudodeterministic Algorithms and the Structure of Probabilistic Time},
  booktitle = {Proceedings of the 53rd Symposium on Theory of Computing (STOC)},
  year      = {2021},
  pages     = {303--316},
  publisher = {ACM},
  doi       = {10.1145/3406325.3451085},
}

@InProceedings{oliveira17pseudodeterministic,
  author    = {Igor Oliveira and Rahul Santhanam},
  title     = {Pseudodeterministic Constructions in Subexponential Time},
  booktitle = {Proceedings of the 49th Symposium on Theory of Computing (STOC)},
  year      = {2017},
  pages     = {665--677},
  publisher = {ACM},
  doi       = {10.1145/3055399.3055500},
}

@InProceedings{goldreich13possibilities,
  author    = {Oded Goldreich and Shafi Goldwasser and Dana Ron},
  title     = {On the Possibilities and Limitations of Pseudodeterministic Algorithms},
  booktitle = {Proceedings of the 4th Innovations in Theoretical Computer Science Conference (ITCS)},
  year      = {2013},
  pages     = {127--138},
  publisher = {ACM},
  doi       = {10.1145/2422436.2422453},
}

@InProceedings{goldwasser21pseudo,
  author    = {Shafi Goldwasser and Russell Impagliazzo and Toniann Pitassi and Rahul Santhanam},
  title     = {On the Pseudo-Deterministic Query Complexity of {NP} Search Problems},
  booktitle = {Proceedings of the 36th Computational Complexity Conference (CCC)},
  year      = {2021},
  pages     = {36:1--36:22},
  publisher = {Schloss Dagstuhl},
  doi       = {10.4230/LIPICS.CCC.2021.36},
}

@InProceedings{goldwasser20pseudo,
  author    = {Shafi Goldwasser and Ofer Grossman and Sidhanth Mohanty and David Woodruff},
  title     = {Pseudo-Deterministic Streaming},
  booktitle = {Proceedings of the 11th Innovations in Theoretical Computer Science Conference (ITCS)},
  year      = {2020},
  pages     = {79:1--79:25},
  publisher = {Schloss Dagstuhl},
  doi       = {10.4230/LIPICS.ITCS.2020.79},
}

@Article{chattopadhyay25pseudo,
  author  = {Arkadev Chattopadhyay and Yogesh Dahiya and Meena Mahajan},
  title   = {Pseudo-Deterministic Query Complexity of Search Problems},
  journal = {Computational Complexity},
  year    = {2025},
  volume  = {34},
  number  = {2},
  pages   = {20},
  doi     = {10.1007/S00037-025-00266-7},
}

@InProceedings{blondal26borsuk,
  author    = {Ari Blondal and Hamed Hatami and Pooya Hatami and Chavdar Lalov and Sivan Tretiak},
  title     = {{B}orsuk--{U}lam and Replicable Learning of Large-Margin Halfspaces},
  booktitle = {Proceedings of the 58th Symposium on Theory of Computing (STOC)},
  year      = {2026},
  pages     = {529--540},
  publisher = {ACM},
  doi       = {10.1145/3798129.3800771},
}

@InProceedings{braverman23lower,
  author    = {Vladimir Braverman and Robert Krauthgamer and Aditya Krishnan and Shay Sapir},
  title     = {Lower Bounds for Pseudo-Deterministic Counting in a Stream},
  booktitle = {Proceedings of the 50th International Colloquium on Automata, Languages, and Programming (ICALP)},
  year      = {2023},
  pages     = {30:1--30:14},
  publisher = {Schloss Dagstuhl},
  doi       = {10.4230/LIPICS.ICALP.2023.30},
}

@InProceedings{goldwasser17bipartite,
  author    = {Shafi Goldwasser and Ofer Grossman},
  title     = {Bipartite Perfect Matching in Pseudo-Deterministic {NC}},
  booktitle = {Proceedings of the 44th International Colloquium on Automata, Languages, and Programming (ICALP)},
  year      = {2017},
  pages     = {87:1--87:13},
  publisher = {Schloss Dagstuhl},
  doi       = {10.4230/LIPICS.ICALP.2017.87},
}

@InProceedings{ghosh21matroid,
  author    = {Sumanta Ghosh and Rohit Gurjar},
  title     = {Matroid Intersection: {A} Pseudo-Deterministic Parallel Reduction from Search to Weighted-Decision},
  booktitle = {Proceedings of the 25th International Conference on Randomization and Computation (RANDOM)},
  year      = {2021},
  pages     = {41:1--41:16},
  publisher = {Schloss Dagstuhl},
  doi       = {10.4230/LIPICS.APPROX/RANDOM.2021.41},
}

@InProceedings{oliveira18pseudo,
  author    = {Igor Oliveira and Rahul Santhanam},
  title     = {Pseudo-Derandomizing Learning and Approximation},
  booktitle = {Proceedings of the 22nd International Conference on Randomization and Computation (RANDOM)},
  year      = {2018},
  pages     = {55:1--55:19},
  publisher = {Schloss Dagstuhl},
  doi       = {10.4230/LIPICS.APPROX-RANDOM.2018.55},
}

@InProceedings{grossman19reproducibility,
  author    = {Ofer Grossman and Yang Liu},
  title     = {Reproducibility and Pseudo-Determinism in Log-Space},
  booktitle = {Proceedings of the 30th Symposium on Discrete Algorithms (SODA)},
  year      = {2019},
  pages     = {606--620},
  publisher = {ACM--SIAM},
  doi       = {10.1137/1.9781611975482.38},
}

@InProceedings{grossman23tight,
  author    = {Ofer Grossman and Meghal Gupta and Mark Sellke},
  title     = {Tight Space Lower Bound for Pseudo-Deterministic Approximate Counting},
  booktitle = {Proceedings of the 64th Symposium on Foundations of Computer Science (FOCS)},
  year      = {2023},
  pages     = {1496--1504},
  publisher = {IEEE},
  doi       = {10.1109/FOCS57990.2023.00091},
}

@TechReport{goemans19doubly,
  author      = {Michel Goemans and Shafi Goldwasser and Dhiraj Holden},
  title       = {Doubly-Efficient Pseudo-Deterministic Proofs},
  institution = {Electronic Colloquium on Computational Complexity (ECCC)},
  year        = {2019},
  number      = {TR19-135},
  url         = {https://eccc.weizmann.ac.il/report/2019/135/},
}

@InProceedings{goldwasser18pseudo,
  author    = {Shafi Goldwasser and Ofer Grossman and Dhiraj Holden},
  title     = {Pseudo-Deterministic Proofs},
  booktitle = {Proceedings of the 9th Innovations in Theoretical Computer Science Conference (ITCS)},
  year      = {2018},
  pages     = {17:1--17:18},
  publisher = {Schloss Dagstuhl},
  doi       = {10.4230/LIPICS.ITCS.2018.17},
}

@InProceedings{dixon18pseudodeterministic,
  author    = {Peter Dixon and Aduri Pavan and N. V. Vinodchandran},
  title     = {On Pseudodeterministic Approximation Algorithms},
  booktitle = {Proceedings of the 43rd International Symposium on Mathematical Foundations of Computer Science (MFCS)},
  year      = {2018},
  pages     = {61:1--61:11},
  publisher = {Schloss Dagstuhl},
  doi       = {10.4230/LIPICS.MFCS.2018.61},
}

@InProceedings{dixon22pseudodeterminism,
  author    = {Peter Dixon and Aduri Pavan and Jason Vander Woude and N. V. Vinodchandran},
  title     = {Pseudodeterminism: {P}romises and Lowerbounds},
  booktitle = {Proceedings of the 54th Symposium on Theory of Computing (STOC)},
  year      = {2022},
  pages     = {1552--1565},
  publisher = {ACM},
  doi       = {10.1145/3519935.3520043},
}

@InProceedings{dixon21complete,
  author    = {Peter Dixon and Aduri Pavan and N. V. Vinodchandran},
  title     = {Complete Problems for Multi-Pseudodeterministic Computations},
  booktitle = {Proceedings of the 12th Innovations in Theoretical Computer Science Conference (ITCS)},
  year      = {2021},
  pages     = {66:1--66:16},
  publisher = {Schloss Dagstuhl},
  doi       = {10.4230/LIPICS.ITCS.2021.66},
}

@InProceedings{aaronson26pseudo,
  author    = {Hugo Aaronson and Tom Gur and Jiawei Li},
  title     = {Pseudo-Deterministic Quantum Algorithms},
  booktitle = {Proceedings of the 53rd International Colloquium on Automata, Languages, and Programming (ICALP)},
  year      = {2026},
  pages     = {3:1--3:22},
  publisher = {Schloss Dagstuhl},
  doi       = {10.4230/LIPICS.ICALP.2026.3},
}

@InCollection{goldreich25multi,
  author    = {Oded Goldreich},
  title     = {Multi-Pseudodeterministic Algorithms},
  booktitle = {Computational Complexity and Local Algorithms},
  publisher = {Springer},
  year      = {2025},
  pages     = {22--43},
  doi       = {10.1007/978-3-031-88946-2_2},
}

@Article{watson16nonnegative,
  author  = {Thomas Watson},
  title   = {Nonnegative Rank vs.\ Binary Rank},
  journal = {Chicago Journal of Theoretical Computer Science},
  year    = {2016},
  volume  = {2016},
  number  = {2},
  pages   = {1--13},
  doi     = {10.4086/cjtcs.2016.002},
}

@InProceedings{huang25min,
  author    = {Mi-Ying (Miryam) Huang and Xinyu Mao and Shuo Wang and Guangxu Yang and Jiapeng Zhang},
  title     = {A Min-Entropy Approach to Multi-Party Communication Lower Bounds},
  booktitle = {Proceedings of the 40th Computational Complexity Conference (CCC)},
  year      = {2025},
  pages     = {33:1--33:29},
  publisher = {Schloss Dagstuhl},
  doi       = {10.4230/LIPICS.CCC.2025.33},
}

@InProceedings{yang24communication,
  author    = {Guangxu Yang and Jiapeng Zhang},
  title     = {Communication Lower Bounds for Collision Problems via Density Increment Arguments},
  booktitle = {Proceedings of the 56th Symposium on Theory of Computing (STOC)},
  year      = {2024},
  pages     = {630--639},
  publisher = {ACM},
  doi       = {10.1145/3618260.3649607},
}

@InProceedings{lovett22lifting,
  author    = {Shachar Lovett and Raghu Meka and Ian Mertz and Toniann Pitassi and Jiapeng Zhang},
  title     = {Lifting with Sunflowers},
  booktitle = {Proceedings of the 13th Innovations in Theoretical Computer Science Conference (ITCS)},
  year      = {2022},
  pages     = {104:1--104:24},
  publisher = {Schloss Dagstuhl},
  doi       = {10.4230/LIPICS.ITCS.2022.104},
}

@Article{raz99separation,
  author  = {Ran Raz and Pierre McKenzie},
  title   = {Separation of the Monotone {NC} Hierarchy},
  journal = {Combinatorica},
  year    = {1999},
  volume  = {19},
  number  = {3},
  pages   = {403--435},
  doi     = {10.1007/S004930050062},
}

@InProceedings{klauck03rectangle,
  author    = {Hartmut Klauck},
  title     = {Rectangle Size Bounds and Threshold Covers in Communication Complexity},
  booktitle = {Proceedings of the 18th Conference on Computational Complexity (CCC)},
  year      = {2003},
  pages     = {118--134},
  publisher = {IEEE},
  doi       = {10.1109/CCC.2003.1214415},
}

@Article{chen20hardness,
  author  = {Lijie Chen},
  title   = {On The Hardness of Approximate and Exact (Bichromatic) Maximum Inner Product},
  journal = {Theory of Computing},
  year    = {2020},
  volume  = {16},
  pages   = {1--50},
  doi     = {10.4086/TOC.2020.V016A004},
}

@Article{gur18non,
  author  = {Tom Gur and Ron Rothblum},
  title   = {Non-Interactive Proofs of Proximity},
  journal = {Computational Complexity},
  year    = {2018},
  volume  = {27},
  number  = {1},
  pages   = {99--207},
  doi     = {10.1007/S00037-016-0136-9},
}

@InProceedings{abboud17distributed,
  author    = {Amir Abboud and Aviad Rubinstein and Ryan Williams},
  title     = {Distributed {PCP} Theorems for Hardness of Approximation in {P}},
  booktitle = {Proceedings of the 58th Symposium on Foundations of Computer Science (FOCS)},
  year      = {2017},
  pages     = {25--36},
  publisher = {IEEE},
  doi       = {10.1109/FOCS.2017.12},
}

@Article{gur15arthur,
  author  = {Tom Gur and Ran Raz},
  title   = {{A}rthur--{M}erlin Streaming Complexity},
  journal = {Information and Computation},
  year    = {2015},
  volume  = {243},
  pages   = {145--165},
  doi     = {10.1016/j.ic.2014.12.011},
}

@Article{chakrabarti14annotations,
  author  = {Amit Chakrabarti and Graham Cormode and Andrew McGregor and Justin Thaler},
  title   = {Annotations in Data Streams},
  journal = {ACM Transactions on Algorithms},
  year    = {2014},
  volume  = {11},
  number  = {1},
  pages   = {7:1--7:30},
  doi     = {10.1145/2636924},
}

@InProceedings{ghosh24new,
  author    = {Prantar Ghosh and Vihan Shah},
  title     = {New Lower Bounds in {M}erlin--{A}rthur Communication and Graph Streaming Verification},
  booktitle = {Proceedings of the 15th Innovations in Theoretical Computer Science Conference (ITCS)},
  year      = {2024},
  pages     = {53:1--53:22},
  publisher = {Schloss Dagstuhl},
  doi       = {10.4230/LIPICS.ITCS.2024.53},
}

@InProceedings{bhadauria25snark,
  author    = {Rishabh Bhadauria and Alexander Block and Prantar Ghosh and Justin Thaler},
  title     = {{SNARK} Lower Bounds via Communication Complexity},
  booktitle = {Proceedings of the 23rd International Conference on Theory of Cryptography (TCC)},
  year      = {2025},
  pages     = {417--451},
  publisher = {Springer},
  doi       = {10.1007/978-3-032-12287-2_15},
}

@Article{babai91nondeterministic,
  author  = {L{\'{a}}szl{\'{o}} Babai and Lance Fortnow and Carsten Lund},
  title   = {Non-Deterministic Exponential Time has Two-Prover Interactive Protocols},
  journal = {Computational Complexity},
  year    = {1991},
  volume  = {1},
  pages   = {3--40},
  doi     = {10.1007/BF01200056},
}

@InProceedings{pitassi23strength,
  author    = {Toniann Pitassi and Morgan Shirley and Adi Shraibman},
  title     = {The Strength of Equality Oracles in Communication},
  booktitle = {Proceedings of the 14th Innovations in Theoretical Computer Science Conference (ITCS)},
  year      = {2023},
  pages     = {89:1--89:19},
  publisher = {Schloss Dagstuhl},
  doi       = {10.4230/LIPICS.ITCS.2023.89},
}

@InProceedings{goos15lower,
  author    = {Mika G{\"{o}}{\"{o}}s},
  title     = {Lower Bounds for Clique vs.\ Independent Set},
  booktitle = {Proceedings of the 56th Symposium on Foundations of Computer Science (FOCS)},
  year      = {2015},
  pages     = {1066--1076},
  publisher = {IEEE},
  doi       = {10.1109/FOCS.2015.69},
}

\end{document}